\documentclass[acmsmall,screen,nonacm]{acmart}
\newif\ifEnableExtend
\EnableExtendfalse

\usepackage[utf8]{inputenc}
\usepackage{url}
\usepackage{color}

\newcommand{\added}[1]{#1}
\usepackage[normalem]{ulem}
\newcommand{\Is}       {:=}
\newcommand{\set}[1]{\left\{ #1\right\}}
\newcommand{\sodass}{\,:\,}
\newcommand{\setGilt}[2]{\left\{ #1\sodass #2\right\}}
\usepackage{amsmath}
\usepackage{amsthm}
\usepackage{xspace}
\usepackage{relsize}
\usepackage[binary-units]{siunitx}
\usepackage{xcolor}

\usepackage{siunitx}

\newcommand{\ie}{i.e.,}
\newcommand{\eg}{e.g.,}
\newcommand{\ea}{et~al.}
\newcommand{\bigO}{\mathcal{O}}

\newcommand{\softO}{\tilde{\mathcal{O}}}

\newcommand{\OPT}{\textsc{Opt}}
\DeclareMathOperator{\poly}{poly}
\DeclareMathOperator{\polylog}{polylog}

\newcommand{\Path}[1]{\left(#1\right)}

\newcommand{\Arc}[2]{(#1, #2)}
\newcommand{\Degree}[1]{\ensuremath{\mathrm{deg}(#1)}}
\newcommand{\InDegree}[1]{\ensuremath{\mathrm{deg}^\mathsmaller{-}(#1)}}
\newcommand{\OutDegree}[1]{\ensuremath{\mathrm{deg}^\mathsmaller{+}(#1)}}

\newcommand{\Queue}[1]{\Code{Q}}

\newcommand{\Tool}[1]{\textsf{#1}}

\newcommand{\SectLabel}[1]{\label{sect:#1}}
\newcommand{\Section}[1]{Section~\ref{sect:#1}}

\newcommand{\Weight}{\ensuremath{w}}
\newcommand{\MaxWeight}{\ensuremath{W}}
\newcommand{\MaxWeightRatio}{\ensuremath{C}}

\newcommand{\BatchSize}{\ensuremath{B}}

\newcommand{\etal}{et~al.\xspace}

\newcommand{\Comment}[1]{\textsl{#1}}

\newcommand{\eps}[0]{{\epsilon}}

\usepackage{wrapfig}
\usepackage{calc}
\newcommand{\erclogowrapped}[1]{%
\setlength\intextsep{0pt}%
\begin{wrapfigure}[3]{r}{#1*\real{1.1}}%
\includegraphics[width=#1]{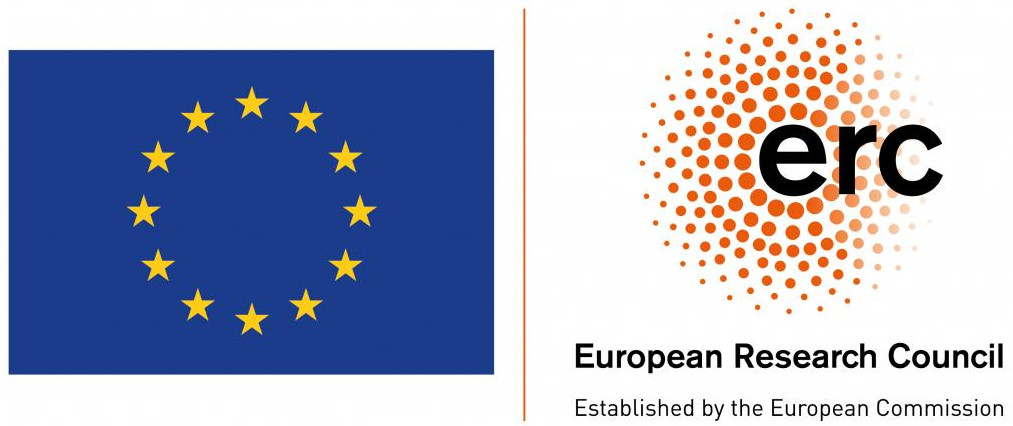}%
\end{wrapfigure}%
}

\setcopyright{acmcopyright}
\copyrightyear{2022}
\acmVolume{}
\acmNumber{}
\acmArticle{}
\acmMonth{12}

\acmYear{2021}
\acmDOI{}
\acmJournal{JEA}
\acmPrice{15.00}
\acmISBN{xx}

\ccsdesc[500]{General and reference~Surveys and overviews}
\ccsdesc[500]{Networks~Network dynamics}
\ccsdesc[500]{Mathematics of computing~Graph algorithms}
\acmBooktitle{ACM Computing Surveys}
\title{Recent Advances in Fully Dynamic Graph Algorithms -- A Quick Reference Guide}
\author{Kathrin Hanauer}
\email{kathrin.hanauer@univie.ac.at}
\affiliation{%
  \institution{University of Vienna, Faculty of Computer Science}
  \streetaddress{Währinger Str. 29}
  \city{Vienna}
  \state{Vienna}
  \country{Austria}
  \postcode{1090}
}\author{Monika Henzinger}
\email{monika.henzinger@univie.ac.at}
\affiliation{%
  \institution{University of Vienna, Faculty of Computer Science}
  \streetaddress{Währinger Str. 29}
  \city{Vienna}
  \state{Vienna}
  \country{Austria}
  \postcode{1090}
}
\author{Christian Schulz}
\email{christian.schulz@informatik.uni-heidelberg.de}
\affiliation{%
  \institution{Heidelberg University}
  \streetaddress{Im Neuenheimer Feld 205}
  \city{Heidelberg}
  \state{Baden-Württemberg}
  \country{Germany}
  \postcode{69120}
}

\date{}

\begin{document}

\begin{abstract}
In recent years, significant advances have been made in the design and analysis of fully dynamic algorithms.  However, these theoretical results have received very little attention from the practical perspective. Few of the algorithms are implemented and tested on real datasets, and their practical potential is far from understood. Here, we present a quick reference guide to recent engineering and theory results in the area of fully dynamic graph algorithms.
\end{abstract}
\maketitle

\section{Introduction}
A (fully) dynamic graph algorithm is a data structure that supports edge insertions, edge deletions, and answers certain queries that are specific to the problem under consideration.
There has been a lot of research on dynamic algorithms for graph problems that are solvable in polynomial time by a static algorithm. The most studied dynamic problems are graph problems such as connectivity, reachability, shortest paths, or matching (see~\cite{DBLP:conf/sofsem/Henzinger18}). Typically, any dynamic algorithm that can handle edge insertions can be used as a static algorithm by starting with an empty graph and inserting all $m$ edges of the static input graph step-by-step. A fundamental question that arises is which problems can be \emph{fully dynamized}, which boils down to the question whether they admit a dynamic algorithm that supports updates in $\bigO(T(m)/m)$ time, where $T(m)$ is the static running time. Thus, for static problems that can be solved in near-linear time, the research community is interested in near-constant time updates. By now, such results have been achieved for a wide range of problems~\cite{DBLP:conf/sofsem/Henzinger18}, which resulted in a rich algorithmic toolbox spanning a wide range of techniques.
However, while there is a large body of theoretical work on efficient dynamic graph algorithms, most of these algorithms were never implemented and   empirically evaluated.
For some classical dynamic algorithms, experimental studies have been performed, such as early works on (all pairs) shortest paths \cite{DBLP:journals/jea/FrigioniINP98,DBLP:journals/talg/DemetrescuI06} or transitive closure~\cite{DBLP:journals/jea/KrommidasZ08} and later contributions for fully dynamic graph clustering~\cite{DBLP:conf/wads/DollHW11} and fully dynamic approximation of betweenness centrality~\cite{DBLP:conf/esa/BergaminiM15}. However, for other fundamental dynamic graph problems, the theoretical algorithmic ideas have received very little attention from the practical perspective. In particular, very little work has been devoted to engineering such algorithms and providing efficient implementations in practice.
Previous surveys on the topic~\cite{zaroliagis2002implementations,DBLP:journals/jea/AlbertsCI97} are more than twenty years old and do not capture the state-of-the-field anymore.
In this work, we aim to survey recent progress in theory as well as
in the empirical evaluation of
fully dynamic graph algorithms and summarize methodologies used to evaluate such algorithms. Moreover, we point to theoretical results that we think have a good potential for practical implementations. Hence, this paper should help an unfamiliar reader by providing most recent references for various problems in fully dynamic graph algorithms. Lastly, there is currently a lack of fully dynamic real-world graphs available online --  most of the instances that can be found to date are insertions-only. Hence, together with this survey we will also start a new open-access graph repository that provides fully dynamic graph instances\footnote{If you have access to fully dynamic instances, we are happy to provide them in our repository. }\footnote{\url{https://DynGraphLab.github.io}}.

We want to point out that there are also various dynamic graph models which we cannot discuss in any depth for space limitations. These are
insertions-only algorithms, deletions-only algorithms, offline dynamic
algorithms%
, algorithms with vertex insertions and deletions, kinetic algorithms%
, temporal algorithms
, algorithms with a limit on the number of allowed queries
, algorithms for the sliding-windows model
, and algorithms for sensitivity problems
(also called emergency planning or fault-tolerant algorithms)
.
We also exclude dynamic algorithms in other models of computation such as distributed algorithms 
 and algorithms in the massively parallel computation (MPC) model.
If the full graph is known at preprocessing time and vertices are ``switched on and off'', this is called the \emph{subgraph model},
whereas \emph{algorithms under failures} deal with the case that vertices or edges are only ``switched off''. We do not discuss these algorithms either.

Note that fully dynamic graph algorithms (according to our definition) are also sometimes called \emph{algorithms for evolving graphs} or \emph{for incremental graphs} or sometimes even \emph{maintaining a graph online}.

\section{Preliminaries}
\paragraph{Basic Definitions}
Let $G = (V, E)$ be a (un)directed graph with vertex set $V$ and edge set
$E$.
Throughout this paper, let $n=|V|$ and $m=|E| \in \bigO(n^2)$.
The \emph{density} of $G$ is $d = \frac{m}{n}$.
In the directed case, an edge $\Arc{u}{v} \in E$ has \emph{tail} $u$ and \emph{head} $v$
and $u$ and $v$ are said to be \emph{adjacent}.
$\Arc{u}{v}$ is said to be an \emph{outgoing} edge or \emph{out-edge} of $u$
and an \emph{incoming} edge or \emph{in-edge} of $v$.
The \emph{outdegree} $\OutDegree{v}$/\emph{indegree}
$\InDegree{v}$/\emph{degree} $\Degree{v}$ of a vertex $v$ is its number of
(out-/in-) edges.
The \emph{out-neighborhood} (\emph{in-neighborhood}) of a vertex
$u$ is the set of all vertices $v$ such that $\Arc{u}{v} \in E$ ($\Arc{v}{u}\in E$).
In the undirected case, $N(v)\Is \setGilt{u}{\set{v,u}\in E}$ denotes the \emph{neighbors} of $v$.
The degree of a vertex $v$ is $\Degree{v}:=|N(v)|$ here.

A graph is \emph{weighted} if there is additionally a function $\Weight: E
\rightarrow \mathbb{R}$ that assigns each edge $e \in E$ some weight $\Weight(e)$.
The maximum weight of any edge in $G$ is denoted as $\MaxWeight$ and
$\MaxWeightRatio$ refers to the ratio of the largest to the smallest edge
weight.
In general, the graph is unweighted unless explicitly mentioned otherwise.

We use $\softO(\cdot)$ to hide polylogarithmic factors.

\paragraph{Dynamic Graphs}
Our focus in this paper are \emph{fully dynamic graphs}, where the set of
vertices is fixed, but edges can be added and removed.
A fully dynamic graph can be seen as a sequence of graphs where two consecutive graphs differ by exactly one edge.
As we only discuss fully dynamic graph algorithms in this survey, we drop the term ``fully'' when it does not lead to a confusion.
For very few problems insertions and deletions of \emph{vertices} instead of \emph{edges} have been studied as well. If so, we will mention them briefly as well.
In case of weighted graphs, an update operation might be supported by the dynamic graph algorithm, namely an update operation that changes the weight of a single edge. If this is the case, we state it explicitly.

Formally, we assume that a dynamic graph algorithm starts with a (possibly empty) initial graph
and may perform some \emph{preprocessing} on it.
It is then presented with a sequence of \emph{updates},
which may in each case either be an \emph{edge insertion}, an \emph{edge
deletion}, or, in case of a weighted graph, a \emph{weight change} of a single
edge.
Unless stated otherwise, (a)~the initial graph is empty and during
preprocessing the data structure is initialized in $\bigO(n)$ time and
(b)~updates are given to the dynamic algorithm one-by-one and it is not able to
look ahead in time.

The sequence of updates may be interspersed with an arbitrary number of
\emph{queries} for an algorithm's current solution.
Depending on the problem setting and the algorithm, the result of a query may
either be an entire solution or just a numeric or boolean value.
A common approach for non-boolean queries is that the algorithm returns a
numeric value, but the entire solution can be obtained optionally in time
linear in the size of the solution.
In some scenarios, the updates are grouped into \emph{batches} of variable size
$\BatchSize$ and an algorithm can process the updates in a batch in arbitrary
order before the next batch or query arrives.
Algorithms that are able to handle batches of updates are also called \emph{batch-dynamic}.
We use \emph{operation} as an umbrella term for both updates and queries.
If the running time for an operation depends on
$n$ or $m$ (or $\MaxWeight$ or $\MaxWeightRatio$, if applicable),
they always refer to the \emph{current} graph unless denoted otherwise.

We do not consider \emph{partially dynamic} algorithms here, \ie{}
algorithms that can either only deal with edge insertions
(also called \emph{incremental} algorithms)
or only with edge deletions
(also called \emph{decremental} algorithms).

\paragraph{Parallel Algorithms}
We consider both sequential and parallel algorithms, where in case of the
latter, the operations are performed by a usually unspecified number of $p$
processors operating in parallel.
In this context, \emph{work} refers to the sum of the running time of all
processors, whereas the \emph{depth} or \emph{span} is the maximum running time
of a single processor if $p$ is infinite.
The depth therefore corresponds to the length of a longest path of operations
that need to be processed sequentially and imposes a lower bound on the
smallest achievable (elapsed) running time.

\paragraph{Conditional Lower Bounds}
There are lower bounds for fully dynamic graph algorithms
based on various popular conjectures initiated by~\cite{Patrascu10,abboud2014popular,henzinger2015unifying}. These lower bounds usually involve three parameters: the preprocessing time $p(m,n)$, the update time $u(m,n)$, and the query time $q(m,n)$. We will use the notation $(p(m,n), u(m,n), q(m,n))$ below to indicate that 
\emph{no algorithm with preprocessing time at most $p(m,n)$ exists that requires update time at most $u(m,n)$ \emph{and} query time at most $q(m,n)$}. 
Note that if the preprocessing time is larger than $p(m,n)$ or if the query time is larger than $q(m,n)$, then it might be possible to achieve an update time better than $u(m,n)$. In the same vein, if the preprocessing time is larger than $p(m,n)$ or if the update time is larger than $u(m,n)$, then it might be possible to achieve a query time better than $q(m,n)$. We will write $\poly(\cdot)$ to denote any running time that is \emph{polynomial} in the parameters.

Any conditional lower bound that is based on the OMv (Online Boolean Matrix-Vector Multiplication) conjecture~\cite{henzinger2015unifying} 
applies to both
the (amortized or worst-case) running time of any fully dynamic algorithm \emph{and also} to the worst-case running time of insertions-only and deletions-only  algorithms. We will not mention this for each problem below and only state the lower bound, except in cases where as a result of the lower bound only algorithms for the insertions-only or deletions-only setting have been studied.

For many of the weighted graph problems the conditional lower bounds continue to hold even if the only update operations are edge weight changes, even if the edge weight can only be increased or decreased by a constant.
More specifically, for any small $\epsilon > 0$ a  conditional lower bound of $(\poly(n), n^{1-\epsilon}, n^{2-\epsilon})$  can be achieved for weighted matching, maximum flow and shortest paths~\cite{DBLP:conf/networking/HenzingerP021}. The same paper presents almost tight upper bounds for these problems.
It also shows a (unconditional) lower bound of $\Omega(\log n)$ on the time per update or query operation for dynamically maintaining a minimum spanning tree. Due to these strong lower bounds we do not discuss edge weight changes in the theoretical results below any further.

\section{Fully Dynamic Graph Algorithms}
In this section, we describe recent efforts in fully dynamic graph algorithms.
We start by describing fundamental problems that we think belong to a basic toolbox of fully dynamic graph algorithms:
strongly connected components,
minimum spanning trees,
cycle detection/topological ordering,
matching,
core decomposition,
subgraph detection,
diameter,
as well as independent sets.
Later on, we discuss problems that are closer to the application side.
To this end we include fully dynamic algorithms for
shortest paths,
maximum flows,
graph clustering,
centrality measures,
and graph partitioning.

\subsection{(Strongly) Connected Components and BFS/DFS Trees}
One of the most fundamental questions on graphs is whether two given vertices
are connected by a path.
In the undirected case, a path connecting two vertices $u$ and $w$ is a
sequence of edges $\mathcal{P} = \Path{\{u,v_0\},\{v_0,v_1\},\dots,\{v_k,w\}}$.
A \emph{connected component} is a maximal set of vertices that are pairwise
connected by a path.
A graph is \emph{connected} if there is exactly one connected component, which
is $V$.
In a directed graph, we say that a vertex $u$ can \emph{reach} a vertex $w$ if
there is a directed path from $u$ to $w$, i.e., a sequence of directed edges
$\mathcal{P} = \Path{(u,v_0),(v_0,v_1),\dots,(v_k,w)}$.
A \emph{strongly connected component} (SCC) is a maximal set of vertices that
can reach each other pairwise.
A directed graph is \emph{strongly connected} if there is just one strongly
connected component, which is $V$.
The \emph{transitive closure} of a graph $G$ is a graph on the same vertex set
with an edge $(u, w) \in V \times V$ if and only if $u$ can reach $w$ in $G$.
Given an undirected graph, we can construct a directed graph from it by
replacing each undirected edge $\{u, w\}$ by a pair of directed edges $(u, w)$
and $(w, u)$ and translate queries of connectedness into reachability queries
on the directed graph.

A breadth-first search (BFS) or depth-first search (DFS) traversal of a
directed or undirected graph defines a rooted tree that consists
of the edges via which a new vertex was discovered.
Apart from connectivity or reachability, BFS and DFS trees can be used to
answer a variety of problems on graphs, such as testing bipartiteness, shortest
paths in the unweighted setting, 2-edge connectivity, or biconnectivity.

\subsubsection{Undirected Graphs (Connectivity)}
\paragraph{Theory Results}
Patrascu and Demaine~\cite{DBLP:journals/siamcomp/PatrascuD06} gave an (unconditional) lower bound of $\Omega(\log n)$ per operation for this problem, improving 
a bound of $\Omega(\log n / \log \log n)$~\cite{DBLP:journals/algorithmica/HenzingerF98}.
The first non-trivial dynamic algorithms for connectivity, and also for 2-edge connectivity, and 2-vertex connectivity~\cite{DBLP:journals/siamcomp/Frederickson97,DBLP:journals/algorithmica/Henzinger95,DBLP:journals/jcss/EppsteinGIS96,DBLP:journals/siamcomp/EppsteinGIS98,DBLP:journals/siamcomp/Henzinger00} took time $\softO(\sqrt{n})$ per operation, including a query which is given two vertices and returns whether they are suitably connected.
Henzinger and King~\cite{DBLP:journals/jacm/HenzingerK99} were the first to give a fully dynamic algorithm with polylogarithmic time per operation for this problem. Their algorithm is, however, randomized. Holm \etal~\cite{DBLP:journals/jacm/HolmLT01} gave the first deterministic fully dynamic algorithm  with polylogarithmic time per operation.
The currently fastest fully dynamic connectivity algorithm takes $\bigO(\log n (\log \log n)^2)$ amortized expected time per operation~\cite{DBLP:conf/soda/HuangHKP17}.
There also is a batch-dynamic parallel algorithm
that answers $k$ queries in $\bigO(k \log(1+n/k))$ expected work and $\bigO(\log n)$ depth
with $\bigO(\log n \log(1+n/\BatchSize))$ expected amortized work per update and $\bigO(\log^3 n)$ depth 
for an average batch size of $\BatchSize$~\cite{DBLP:conf/spaa/AcarABD19}.

The fully dynamic connectivity problem can be reduced to the maintenance of a
spanning forest, using, \eg{} dynamic
trees~\cite{DBLP:conf/stoc/SleatorT81,DBLP:conf/esa/AcarABDW20} or Euler tour
trees~\cite{HenzingerKing95,DBLP:conf/alenex/TsengDB19} (see also
\Section{mst}), for the components.
If the graph is a forest, updates and queries can be processed in amortized
$\bigO(\log n)$ time, whereas the theoretically fastest
algorithms~\cite{DBLP:conf/soda/KapronKM13} to date for general graphs have
polylogarithmic worst-case update time and $\bigO(\log n/\log\log n)$ worst-case query time, the latter matching the lower
bound~\cite{DBLP:journals/algorithmica/HenzingerF98,DBLP:journals/tcs/MiltersenSVT94}.
The key challenge on general graphs is to determine whether the deletion of an
edge of the spanning forest disconnects the component or whether a replacement
edge can be found.
There are also fully dynamic algorithms for more refined notions of connectivity:
Two-edge connectivity~\cite{DBLP:journals/jacm/HenzingerK99,HolmLT98} and two-vertex connectivity~\cite{HolmLT98} can also be maintained in polylogarithmic time per operation. See~\cite{DBLP:reference/algo/Italiano16b} for a survey on that topic.

\paragraph{Experimental Results}
Building on an earlier study by Alberts
\etal{}~\cite{DBLP:journals/jea/AlbertsCI97},
Iyer~\etal{}~\cite{DBLP:journals/jea/IyerKRT01} experimentally compared the
Euler tour tree-based algorithms by Henzinger and King~\cite{HenzingerKing95}
and Holm~\etal{}~\cite{HolmLT98} to each other as well as several heuristics to
achieve speedups in both candidates.
The instances used in the evaluation were random graphs with random edge
insertions and deletions, random graphs where a fixed set of edges appear and
disappear dynamically, graphs consisting of cliques of equal size plus a set of
inter-clique edges, where only the latter are inserted and deleted, as well as
specially crafted worst-case instances for the algorithms.
The authors showed that the running time of both algorithms can be improved
distinctly via heuristics; in particular a sampling approach to replace deleted
tree edges has proven to be successful.
The experimental running time of both algorithms was comparable, but with the
heuristics, the algorithm by Holm~\etal{}~\cite{HolmLT98} performed better.

Baswana \ea{}~\cite{DBLP:journals/siamcomp/BaswanaCC019} gave the first
algorithm for maintaining an undirected DFS tree with $o(m)$ update time and
showed a conditional lower bound of $\Omega(n)$ on the update time in case of
vertex updates and, if the tree is maintained explicitly,
an unconditional lower bound of $\Omega(n)$ under edge updates.
Their algorithm has a preprocessing time of $\bigO(m\log n)$, a worst-case
update time of $\bigO(\sqrt{mn}\log^{2.5} n)$, and uses $\bigO(m \log^2 n)$
bits.
Nakamura and Sadakane~\cite{DBLP:journals/algorithms/NakamuraS19} improved the
update time by $\polylog n$ factors and the space required to $\bigO(m \log
n)$.
Recently, Baswana \ea{}~\cite{DBLP:conf/mfcs/BaswanaGT19} further reduced the
update time down to $\bigO(\sqrt{mn\log n})$.
A parallel algorithm that uses $m$ processors and $\bigO(\polylog n)$ update
time was given by Khan~\cite{DBLP:journals/topc/Khan19}.
To the best of our knowledge, experimental evaluations have only been conducted
to date with algorithms designed for the \emph{incremental} setting, but not
for fully-dynamic algorithms.

No experimental studies on dynamically maintaining BFS trees are known to us.
%
%
\ifEnableExtend
2001 ``An Experimental Study of Polylogarithmic, Fully Dynamic, Connectivity Algorithms''~\cite{DBLP:journals/jea/IyerKRT01}
\fi
\subsubsection{Directed Graphs (Reachability, Strong Connectivity, Transitive Closure)}
\paragraph{Theory Results}
There is a conditional lower bound of
$(\poly(n), m^{1/2-\epsilon}, m^{1-\epsilon})$ for any small constant $\epsilon >0$ based on the
OMv conjecture. This bound even holds for the $s$-$t$ reachability problem, where
both $s$ and $t$ are fixed for all queries.
The currently fastest algorithms for transitive closure are three Monte Carlo algorithms with one-sided error:
Two by Sankowski~\cite{DBLP:conf/focs/Sankowski04} with $\bigO(1)$ or
$\bigO(n^{0.58})$ worst-case query time and $\bigO(n^2)$ or $\bigO(n^{1.58})$
worst-case update time, respectively, and one by van den Brand, Nanongkai, and
Saranurak~\cite{DBLP:conf/focs/BrandNS19} with $\bigO(n^{1.407})$ worst-case
update and worst-case query time. Both algorithms can only answer a reachability query, but if a vertex $x$ can reach a vertex $y$, the algorithms cannot actually output a path from $x$ to $y$.
There exists a conditional lower bound
based on a variant of the OMv conjecture that shows that these
running times are optimal~\cite{DBLP:conf/focs/BrandNS19}.
Moreover, there are two deterministic, combinatorial algorithms:
Roditty's algorithm with constant query time and $\bigO(n^2)$ amortized update
time~\cite{DBLP:journals/talg/Roditty08}, as well as one by Roditty and Zwick~\cite{DBLP:journals/siamcomp/RodittyZ16} with an
improved $\bigO(m+n\log n)$ amortized update time at the expense of $\bigO(n)$
worst-case query time. The former algorithm can also return a directed path from a given vertex $x$ to a given vertex $y$ if such a path exists in time linear in its length, and also supports the update of a vertex, not only of an edge in time $\bigO(n^2)$.

\paragraph{Experimental Results for Single-Source Reachability}
The single-source reachability problem, where a vertex $s$ is fixed and
reachability from $s$ to any other vertex $v$ is the subject of queries,
was first studied experimentally by Hanauer \etal~%
\cite{DBLP:conf/alenex/HanauerH020
}.
Here, two relatively straightforward algorithms, called \Tool{SI} and
\Tool{SES},
showed outstanding performance in practice
on both random graphs as well as on real-world instances
in comparison to
the static baseline algorithms breadth-first and depth-first search,
which were up to this point the recommended
choice to maintain reachability in a dynamic
setting~\cite{DBLP:journals/jea/FrigioniMNZ01,DBLP:journals/jea/KrommidasZ08}.
\Tool{SI} maintains an arbitrary reachability tree which is re-constructed via
a combined forward and backward breadth-first search traversal on edge
deletions if necessary and is especially fast if insertions predominate,
which can be handled in $\bigO(n+m)$ time.
By contrast, it may take up to $\bigO(nm)$ time for a single edge removal.
\Tool{SES} is an extension and simplification of Even-Shiloach
trees~\cite{ES81}, which originally only handle edge deletions.
Its strength are hence instances with many deletions.
As a plus, it is able to deliver not just any path as a witness for
reachability, but even the shortest path (with respect to the number of edges).
Furthermore, it internally maintains a BFS tree, which makes it viable
also for numerous other applications, see above.
Its worst-case update time is $\bigO(n+m)$, and, like \Tool{SI}, it answers
queries in constant time.
One key ingredient for the superior performance of both algorithms in practice
are carefully chosen criteria for an abortion of the re-construction of their
data structures and their re-building from
scratch. 

\paragraph{Experimental Results for Transitive Closure}
Frigioni~\ea{}~\cite{DBLP:journals/jea/FrigioniMNZ01} and later Krommidas and
Zaroliagis~\cite{DBLP:journals/jea/KrommidasZ08} empirically studied the
performance of an extensive number of algorithms for transitive closure,
including those mentioned above.
They also developed various extensions and variations and compared them not
only to each other, but also to static, so-called ``simple-minded'' algorithms
such as breadth-first and depth-first search.
Their evaluation included random Erd\H{o}s-Renyí graphs, specially constructed
hard instances, as well as two instances based on real-world graphs.
It showed that the ``simple-minded'' algorithms could outperform the dynamic
ones distinctly and up to several factors, unless the query ratio was more than
\SI{65}{\percent} or the instances were dense random graphs.

In a follow-up experimental study by Hanauer \etal~%
\cite{DBLP:conf/wea/HanauerH020},
the authors showed that the ``simple-minded'' algorithms can be outperformed by
several orders of magnitude in practice again on both random and real-world
dynamic graphs by maintaining multiple instances of their single-source
reachability algorithms simultaneously.
To query the transitive closure of a graph, a number of so-called ``supportive
vertices'', for which both in- and out-reachability trees are maintained
explicitly, can be picked either once or periodically anew and then be used to
answer both positive and negative reachability queries between a number of
pairs of vertices decisively in constant time. 
The fallback routine can be a simple static graph traversal and therefore
be relatively expensive:
With a random initial choice of supportive vertices and no periodic renewals,
this approach has been shown to answer a great majority of reachability queries
on both random and real-world instances in constant time already if the number
of supportive vertices is very small, i.e., two or three.

These experimental studies clearly show the limitations of worst-case analysis:
All implemented algorithms are fully dynamic with at least linear worst-case
running time per operation and, thus, all perform ``(very) poor''
in the worst case.
Still on all graphs used in the study the relatively simple new algorithms
clearly outperformed the algorithms used in previous studies.
\ifEnableExtend

2001 ``An Experimental Study of Dynamic Algorithms for Transitive Closure''~\cite{DBLP:journals/jea/FrigioniMNZ01} \\
2008 ``An experimental study of algorithms for fully dynamic transitive closure''~\cite{DBLP:journals/jea/KrommidasZ08} \\
2013 ``DAGGER: A Scalable Index for Reachability Queries in Large Dynamic Graphs''~\cite{DBLP:journals/corr/abs-1301-0977} \Comment{arxiv only} \\
2020 ``Fully Dynamic Single-Source Reachability in Practice: An Experimental Study''~\cite{DBLP:conf/alenex/HanauerH020} \\
2020 ``Faster Fully Dynamic Transitive Closure in Practice''~\cite{DBLP:conf/wea/HanauerH020}
\fi

\paragraph{Experimental Results for DFS/BFS Trees}
Yang \ea{}~\cite{DBLP:journals/pvldb/YangWQ0WL19} were the first to give a fully dynamic
algorithm for maintaining a DFS tree in a directed graph along with several
optimizations to achieve speedups in practice.
In an experimental evaluation on twelve real-world instances, they showed that
the optimized version of their algorithm can handle edge insertions and deletions
within few seconds on average for instances with millions of vertices.

With regard to BFS trees,
the already mentioned \Tool{SES} algorithm~\cite{DBLP:conf/alenex/HanauerH020}
is the only fully dynamic algorithm we are aware of that maintains a BFS tree
on a directed graph.

\subsection{Minimum Weight Spanning Trees}\SectLabel{mst}
A minimum weight spanning tree (MST) of a connected graph is a subset of the edges such that all nodes are connected via the edges in the subset, the induced subgraph has no cycles and, lastly, has the minimum total weight among all possible subsets fulfilling the first two properties.

\paragraph{Theory Results}
The lower bound of $\Omega(\log n)$~\cite{DBLP:journals/siamcomp/PatrascuD06} on the time per operation for connectivity trivially extends to maintaining the weight of a minimum spanning tree.
Holm \etal~\cite{DBLP:journals/jacm/HolmLT01} gave the first fully dynamic algorithm with polylogarithmic time per operation for this problem. It was later slightly improved to $\bigO(\log^4 n/\log \log n)$ time per operation
\cite{DBLP:conf/esa/HolmRW15}.

\paragraph{Experimental Results}
Amato \etal~\cite{DBLP:conf/soda/AmatoCI97} presented the first experimental study of dynamic minimum spanning tree algorithms. In particular, the authors implemented different versions of Frederickson's algorithm~\cite{DBLP:journals/siamcomp/Frederickson85} which uses partitions and topology trees.
The algorithms have been adapted with sparsification techniques to improve their performance. 
The update running times of these algorithms range from $\bigO(m^{2/3})$ to $\bigO(m^{1/2})$. The authors further presented a variant of Frederickson's algorithm that is significantly faster than all other implementations of this algorithm. However, the authors also proposed a simple adaption of a partially dynamic data structure of Kruskal's algorithm that was the fastest implementation on random inputs.
Later, Cattaneo \etal~\cite{DBLP:conf/alenex/CattaneoFPI02,DBLP:journals/dam/CattaneoFPI10} presented an experimental study on several algorithms for the problem. The authors presented an efficient implementation of the algorithm of Holm \etal~\cite{DBLP:journals/jacm/HolmLT01}, proposed new simple algorithms for dynamic MST that are not as asymptotically efficient as the algorithm by Holm \etal but seem to be fast in practice, and lastly compared their algorithms with the results of Amato \etal~\cite{DBLP:conf/soda/AmatoCI97}. The algorithm by Holm \etal uses a clever refinement of a technique by Henzinger and King~\cite{henzinger1997maintaining} for developing fully dynamic algorithms starting from the deletions-only case. One outcome of their experiments is that simple algorithms outperform the theoretically more refined algorithms on random and worst-case networks. On the other hand, on $k$-clique inputs, \ie{} graphs that contain $k$ cliques of size $c$ plus $2k$ randomly chosen inter-clique edges, the implementation of the algorithm by Holm \etal outperformed the simpler algorithms.

Tarjan and Werneck~\cite{DBLP:journals/jea/TarjanW09} performed experiments for several variants of dynamic trees data structure.
The evaluated data structures have been used by Ribero and Toso~\cite{DBLP:conf/wea/RibeiroT07}, who focused on the case of changing weights, \ie{} the edges of the graph are constant, but the edge weights can change dynamically. Ribero and Toso also proposed and used a new data structure for dynamic tree representation called DRD-trees. In their algorithm the dynamic tree data structure is used to speed up connectivity queries that check whether two vertices belong to different subtrees. More generally, Ribero and Toso compared different types of data structures to do this task. In particular, the authors used the dynamic tree data structures that have been evaluated by Tarjan and Werneck~\cite{DBLP:journals/jea/TarjanW09}. The experimental evaluation demonstrated that the new structure reduces the computation time observed for the algorithm of Cattaneo \etal~\cite{DBLP:conf/alenex/CattaneoFPI02}, and at the same time yielded the fastest algorithms in the experiments. 

\ifEnableExtend
        1997 ``Experimental Analysis of Dynamic Minimum Spanning Tree Algorithms''~\cite{DBLP:conf/soda/AmatoCI97} \\
        1997 ``Maintaining minimum spanning trees in dynamic graphs''~\cite{henzinger1997maintaining} \\
2002 ``Maintaining Dynamic Minimum Spanning Trees: An Experimental Study''~\cite{DBLP:conf/alenex/CattaneoFPI02}\\
2007 ``Experimental Analysis of Algorithms for Updating Minimum Spanning Trees on Graphs Subject to Changes on Edge Weights''~\cite{DBLP:conf/wea/RibeiroT07} \\
2010 ``Maintaining dynamic minimum spanning trees: An experimental study''~\cite{DBLP:journals/dam/CattaneoFPI10} \\
2016 ``Fully Dynamic Minimum Spanning Trees''~\cite{DBLP:reference/algo/Italiano16d} 
\fi{}

\subsection{Cycle Detection and Topological Ordering}
A cycle in a (directed) graph $G=(V,E)$ is a non-empty path $\mathcal{P} = \Path{v_1, \ldots, v_k = v_1}$ such that $(v_i, v_{i+1}) \in E$.
A topological ordering of a directed graph is a linear ordering of its vertices from 1 to $n$ such that for every directed edge $(u,v)$ from vertex $u$ to vertex $v$, $u$ is ordered before $v$. 
In the static case, one can use a depth-first search (DFS) to compute a topological ordering of a directed acyclic graph or to check if a (un)directed graph contains a cycle. 

\paragraph{Theory Results}
Let $\epsilon > 0$ be any small constant.
Based on the OMv conjecture~\cite{henzinger2015unifying} it is straightforward to construct a lower bound of $(\poly(n), m^{1/2-\epsilon}, m^{1-\epsilon})$  for the (amortized or worst-case) running time of any fully dynamic algorithm that detects whether the graph contains any cycle. As any algorithm for topological ordering can be used to decide whether a graph contains a cycle, this lower bound also applies to any fully dynamic  topological ordering algorithm.
Via dynamic matrix inverse one can maintain fully dynamic directed cycle detection in $\bigO(n^{1.407})$~\cite{DBLP:conf/focs/BrandNS19}, which is conditionally optimal based on a variant of the OMv conjecture.

\paragraph{Experimental Results}
   Pearce and Kelly~\cite{DBLP:conf/wea/PearceK04,DBLP:journals/jea/PearceK06} were the first to evaluate algorithms for topological ordering in the presence of edge insertions and deletions. In their work, the authors compared three algorithms that can deal with the online topological ordering problem. 
More precisely, the authors implemented the algorithms by Marchetti-Spaccamela \etal{}~\cite{DBLP:journals/ipl/Marchetti-SpaccamelaNR96} and Alpern \etal~\cite{DBLP:conf/soda/AlpernHRSZ90} as well as a newly developed algorithm.
   Their new algorithm is the one that performed best in their experiments. 
The algorithm maintains a node-to-index map, called $n2i$, that maps each vertex to a unique integer in $\{1 \ldots n\}$ and ensures that for any edge $(u,v)$ in $G$, it holds $n2i[u] < n2i[v]$.
When an insertion $(u,v)$ invalidates the topological ordering, affected nodes are updated. The set of affected nodes are identified using a forward DFS from $v$ and backward DFS from $u$. The two sets are then separately sorted into increasing topological order and afterwards a remapping to the available indices is performed. 
The algorithm by Marchetti-Spaccamela \etal{}~\cite{DBLP:journals/ipl/Marchetti-SpaccamelaNR96} is quite similar to the algorithm by Pearce and Kelly. However, it only maintains the forward set of affected nodes and obtains a correct solution by shifting the affected nodes up in the ordering (putting them after $u$).
Alpern \etal~\cite{DBLP:conf/soda/AlpernHRSZ90} used a data structure to create new priorities between existing ones in constant worst-case time. 
The result by Pearce and Kelly has later been applied to online cycle detection and difference propagation in pointer analysis by Pearce \etal~\cite{DBLP:journals/sqj/PearceKH04}. Furthermore, Pearce and Kelly~\cite{Pearce2010ABA} later extended their algorithm to be able to provide more efficient batch updates.

\ifEnableExtend
2004 ``A Dynamic Algorithm for Topologically Sorting Directed Acyclic Graphs''~\cite{DBLP:conf/wea/PearceK04} \\
2004 ``Online Cycle Detection and Difference Propagation: Applications to Pointer Analysis''~\cite{DBLP:journals/sqj/PearceKH04} \\
2006 ``A dynamic topological sort algorithm for directed acyclic graphs''~\cite{DBLP:journals/jea/PearceK06} \\
2007/2010 ``A dynamic batch algorithm for maintaining a topological order'~\cite{pearce2007dynamic} \\
\fi{}

\subsection{(Weighted) Matching}
The matching problem is one of the most prominently studied combinatorial graph problems having a variety of practical applications. 
A matching $\mathcal{M}$ of a graph $G=(V,E)$ is a subset of edges such that no two
elements of $\mathcal{M}$ have a common end point. Many applications require 
matchings with certain properties, like being maximal (no edge can
be added to $\mathcal{M}$ without violating the matching property) or having maximum
cardinality.

\subsubsection{Cardinality Matching}
\paragraph{Theory Results}
There is a conditional lower bound of
$(\poly(n), m^{1/2-\delta}, m^{1-\delta})$ (for any small constant $\delta >0$) for the size of the maximum cardinality matching based on the OMv conjecture~\cite{henzinger2015unifying}. 
Of course, maintaining an actual maximum matching is only harder than maintaining the size of a maximum matching. Thus upper bounds have mostly focused on approximately maximum matching. 
However, also here we have to distinguish between (a) algorithms that maintain the \emph{size} of an approximately maximum matching and (b) algorithms that maintain an \emph{approximately~maximum~matching}. The algorithms below either assume that the algorithm is started on an empty graph or, in the case of the exact algorithms with non-empty initial graphs, they use $\bigO(n^{\omega})$ preprocessing~time.

Note that the query time of all
type-(a) algorithms is constant. As type-(b) algorithms maintain an actual matching the queries they support are more flexible: they can output the matching in time linear in its size, answer queries whether a given edge belongs to the matching in constant time, and output the size of the matching in constant time.

(a) Improving Sankowski's $\bigO(n^{1.495})$ update time bound~\cite{Sankowski07},
van den Brand \etal~\cite{DBLP:conf/focs/BrandNS19} maintain the exact size of a maximum matching in $\bigO(n^{1.407})$ update time.  To maintain the approximate size of the maximum matching, dynamic algorithms use the duality of maximum matching and vertex cover and maintain instead a $(2+\eps)$-approximate vertex cover. This line of work lead to a sequence of papers~\cite{IvkovicL93,BhattacharyaHI18,Bhattacharya2016,DBLP:journals/algorithmica/BhattacharyaCH20}, resulting finally in a deterministic $(2+\epsilon)$-approximation algorithm that maintains a hierarchical graph decomposition with $\bigO(1/\epsilon^2)$ amortized update time~\cite{DBLP:conf/soda/BhattacharyaK19}.  The algorithm can be turned into an algorithm with worst-case $\bigO(\log^3 n)$ time per update~\cite{Bhattacharya2017b}.

(b)
One can trivially maintain a maximal  matching in $\bigO(n)$ update time by resolving all trivial augmenting paths, \ie{} cycle-free paths that start and end on a unmatched vertex and where edges from $\mathcal{M}$ alternate with edges from $E \setminus \mathcal{M}$, of length one. As any maximal matching is a 2-approximation of a maximum matching, this leads to a 2-approximation algorithm.
Onak and Rubinfeld~\cite{OnakRubinfeld10} presented a randomized algorithm for maintaining an $\bigO(1)$-approximate matching with $\bigO(\log^2 n)$ expected amortized time per edge update. 
Baswana, Gupta, and Sen~\cite{BaswanaGS15} gave an elegant algorithm that maintains a  \emph{maximal} matching with amortized update time $\bigO(\log n)$. It is based on a hierarchical graph decomposition and was subsequently improved by Solomon to amortized constant expected update time~\cite{Solomon16}.

For worst-case bounds, the best results are a $(1+\eps)$-approximation in $\bigO(\sqrt{m}/\epsilon)$ update time  by Gupta and Peng~\cite{GuptaP13} (see~\cite{NeimanS16} for a 3/2-approximation in the same time),
a $(3/2+\epsilon)$-approximation in $\bigO(m^{1/4}/\epsilon^{2.5})$ time
by Bernstein and Stein~\cite{Bernstein2016a}, and a $(2+\epsilon)$-approximation in $\bigO(\polylog n)$ time by
 Charikar and Solomon~\cite{CharikarS18} and Arar et al.~\cite{ArarCCSW18}.
There exists also algorithms that achieve
various tiny improvements over the approximation factor of 2~\cite{DBLP:conf/soda/BehnezhadLM20,DBLP:conf/innovations/RoghaniSW22}
with (small) polynomial update time.
Recently, Bernstein \etal~\cite{DBLP:conf/soda/BernsteinFH19} improved the maximal matching algorithm of Baswana \etal~\cite{BaswanaGS15}
to $\bigO(\log^5 n)$ worst-case time with high probability.

\paragraph{Experimental Results}
Despite this variety of different algorithms, to the best of our knowledge, there have been only  limited efforts  so far to engineer and evaluate these algorithms on real-world instances. 
Henzinger \etal~\cite{DBLP:conf/esa/Henzinger0P020} initiated the empirical evaluation of algorithms for this problem in practice.
To this end, the authors evaluated several dynamic maximal matching algorithms as well as an algorithm that is able to maintain the maximum matching. 
They implemented the algorithm by Baswana, Gupta and Sen~\cite{BaswanaGS15}, which performs edge updates in  $\bigO(\sqrt{n})$ time and maintains a 2-approximate maximum matching, the algorithm of Neiman and Solomon~\cite{NeimanS16}, which takes $\bigO(\sqrt{m})$ time to maintain a 3/2-approximate maximum matching, as well as two novel dynamic algorithms, namely a random walk-based algorithm as well as a dynamic algorithm that searches for augmenting paths using a (depth-bounded) blossom algorithm. 
Their experiments indicate that an optimum matching can be maintained dynamically more than an order of magnitude faster
than the naive algorithm that recomputes maximum matchings from scratch. Second, all non-optimum dynamic algorithms that have been considered in that work were able to maintain near-optimum matchings in practice while being multiple orders of magnitudes faster than the naive exact dynamic algorithm. The study concludes that in practice an extended random walk-based algorithms is the method of choice.  

\subsubsection{Weighted Matching}
\paragraph{Theory Results}
For the \emph{weighted} dynamic matching problem, Anand \etal~\cite{DBLP:conf/fsttcs/AnandBGS12} proposed an algorithm that can maintain an \num{4.911}-approximate dynamic maximum weight matching that runs in amortized $\bigO(\log n \log \MaxWeightRatio)$ time where $\MaxWeightRatio$ is the ratio of the weight of the highest weight edge to the weight of the smallest weight edge. Furthermore, a sequence~\cite{DBLP:journals/iandc/BhattacharyaHI18,DBLP:conf/stoc/AbboudA0PS19,DBLP:journals/algorithmica/BhattacharyaCH20,DBLP:journals/corr/abs-2002-11171,DBLP:conf/focs/BhattacharyaHN19}
of work on fully dynamic set cover
resulted in
$(1 + \eps)$-approximate weighted dynamic matching algorithms, with $\bigO(1/\eps^3 + (1/\eps^2) \log \MaxWeightRatio)$ amortized and $\bigO((1/\eps^3)\log^2(\MaxWeightRatio n))$ worst-case
time per operation based on various hierarchical hypergraph decompositions.
Gupta and Peng~\cite{GuptaP13} maintain a $(1+\epsilon)$-approximation under edge insertions/deletions that runs in time $\bigO(\sqrt{m}\epsilon^{-2-O(1/\epsilon)}\log \MaxWeight)$ time per update, if edge weights are in between $1$ and $\MaxWeight$.

Their result is based on rerunning a static algorithm from time to time, a trimming routine that trims the graph to a smaller equivalent graph whenever possible and in the weighted case, a partition of the weights of the edges into intervals of geometrically increasing size.
Stubbs and Williams~\cite{DBLP:conf/innovations/StubbsW17} presented metatheorems for dynamic weighted matching. Here, the authors reduced the dynamic maximum weight matching problem to the dynamic maximum cardinality matching problem in which the graph is unweighted. The authors proved that using this reduction, if there is an $\alpha$-approximation for maximum cardinality matching with update time $T$ in an unweighted graph, then there is also a $(2+\epsilon)\alpha$-approximation for maximum weight matching with update time $\bigO(\frac{T}{\epsilon^2}\log^{2} \MaxWeight)$. Their basic idea is an extension of the algorithm of Crouch and Stubbs~\cite{DBLP:conf/approx/CrouchS14} who tackled the problem in the streaming model. Here, the reduction is to take matchings from weight-threshold based subgraphs of the dynamic graph, \ie{} the algorithm maintains maximal matchings in $\log C$ subgraphs, where subgraph $i$ contains all edges having weight at least $(1+\epsilon)^i$. The resulting matchings are then greedily merged together by considering the matched edges in descending order of $i$ (heaviest edges first).

\paragraph{Experimental Results}
Recently, the approach by Stubbs and Williams has been evaluated experimentally and has been compared against a new random walk-based approach~\cite{DBLP:journals/corr/abs-2104-13098} which gives a $(1+\epsilon)$ approximation w.h.p.. When inserting or deleting an edge, the random walk-based approach finds random simple paths (using random walks) and solves those paths using dynamic programming to improve the maintained matching. In practice, the random walk-based approach outperforms the approach by Stubbs and Williams significantly.

\ifEnableExtend

2012 ``Maintaining Approximate Maximum Weighted Matching in Fully Dynamic Graphs''~\cite{DBLP:conf/fsttcs/AnandBGS12} \\
2013 ``Fully Dynamic (1+$\epsilon$)-Approximate Matchings''~\cite{DBLP:conf/focs/GuptaP13} \\
2014 ``Improved Streaming Algorithms for Weighted Matching, via Unweighted Matching''~\cite{DBLP:conf/approx/CrouchS14} \\
2017 ``Metatheorems for Dynamic Weighted Matching''~\cite{DBLP:conf/innovations/StubbsW17} \\
2019 ``Fully Dynamic Graph Algorithms Inspired by Distributed Computing: Deterministic Maximal Matching and Edge Coloring in Sublinear Update-Time''~\cite{DBLP:journals/jea/BarenboimM19}\\
2020 ``Dynamic Matching in Practice''~\cite{DBLP:conf/esa/Henzinger0P020} \\
2020 ``Lazy or eager dynamic matching may not be fast''~\cite{kashyop2020105982} \Comment{lower bounds} 
\fi{}

\subsection{$k$-Core Decomposition}
A $k$-core of a graph is a maximal connected subgraph in which all vertices have degree at least $k$.
The core number of a vertex is the largest value $k$ such that the vertex is still contained in the $k$-core.
The $k$-core decomposition problem is to compute the core number of every node in the graph.
A related problem is that of finding a \emph{densest subgraph}, which is an
induced subgraph that has maximum density.

\paragraph{Theory Results}
It is well-known that a $k$-core decomposition can be computed in linear time
for a static graph by repeatedly removing vertices of degree less than $k$ from
the graph.
A $2$-approximation for the densest subgraph problem can be obtained by
replacing $k$ with the minimum vertex degree in each step until it is empty and
returning the induced subgraph on the set of vertices $V_i$ that has the
largest density, where $V_0 = V$ and $V_i$ is the set of vertices after step
$i$~\cite{DBLP:conf/approx/Charikar00}.

The problem of maintaining the $k$-core decomposition in a fully dynamic graph has not received much attention by the theoretical computer science community:
 Sun \etal~\cite{10.1145/3385416} showed that the insertion and deletion of a single edge can change the core value of all vertices. They also gave a $(4+\eps)$-approximate fully dynamic algorithm that maintains core values with polylogarithmic running time.
The algorithm can be implemented in time $\bigO(\log^2 n)$ in graphs using the algorithm of Bhattacharya \etal{}~\cite{DBLP:conf/stoc/BhattacharyaHNT15}. It dynamically maintains $\bigO(\log_{(1+\eps)} n)$ many $(\alpha,\beta)$-decompositions of the graph, one
for each $\beta$-value that is a power of $(1+\eps)$ between $1$ and $(1+\eps)n$.

For a subset $Z \subseteq V$ let $\mathrm{deg}_{Z}(v)$ denote the number of neighbors of $v$ in $Z$.
An
$(\alpha,\beta)$-\emph{decomposition} of a graph $G=(V,E)$~\cite{DBLP:conf/stoc/BhattacharyaHNT15} for $\alpha \geq 1, \beta \geq 0$ is a decomposition
$Z_1, \dots , Z_L$ of $V$ into $L := 1+\lceil{(1+\eps)} \log n\rceil$ levels such that $Z_{i+1} \subseteq Z_{i}$ for all $1 \leq i < L$, $Z_1 = V$, and the following invariants are maintained: (1) All vertices $v$ on level $Z_i$  with $\mathrm{deg}_{Z_i}(v) > \alpha \beta$ belong to $Z_{i+1}$ and (2) all vertices $v$ on level $Z_i$  with $\mathrm{deg}_{Z_i}(v) < \beta$ do not belong to $Z_{i+1}$.
An $(\alpha,\beta)$-decomposition is hence an approximate version of the procedure for the static setting described above,
where in each step all vertices with degree less than $\beta$ plus some with degree less than $\alpha\beta$ are removed.
The technique can also be used to maintain a $2\alpha(1 + \eps)^2$-approximate densest subgraph~\cite{DBLP:conf/stoc/BhattacharyaHNT15}.

There are no further lower bounds, neither conditional nor unconditional, and no faster algorithms known for maintaining an approximate $k$-core decomposition.

\paragraph{Experimental Results}
Miorandi and De Pellegrini~\cite{DBLP:conf/wiopt/MiorandiP10} proposed two methods to rank nodes according to their $k$-core number in fully dynamic networks and compare them experimentally. The focus of their work is to identify the most influential spreaders in complex dynamic networks.
Li \etal~\cite{DBLP:journals/tkde/LiYM14} used a filtering method to only update nodes whose core number is affected by the network update. More precisely, the authors showed that nodes that need to be updated must be connected via a path to the endpoints of the inserted/removed edge and the core number must be equal to the smaller core number of the endpoints. Moreover, the authors presented efficient algorithms to identify such nodes as well as additional techniques to reduce the size of the nodes that need updates. 
Similarly, Sariy\"uce~\etal~\cite{DBLP:journals/vldb/SariyuceGJWC16} proposed the $k$-core algorithm \Tool{TRAVERSAL} and gave additional rules to prune the size of  the subgraphs that are guaranteed to contain the vertices whose $k$-core number can have changed. Note that this algorithm can have a high variation in running time for the update operations depending on the size of the affected subgraphs.
Zhang \etal~\cite{DBLP:conf/icde/ZhangYZQ17} noted that due to this reason it can be impractical to process updates one by one and introduced the $k$-order concept which can reduce the cost of the update operations. A $k$-order is defined as follows: a node $u$ is ordered before $v$ in the $k$-order if $u$ has a smaller core number than $v$ or when the vertices have the same core number, if the linear time algorithm to compute the core decomposition would remove $u$ before $v$. 
A recent result by Sun \etal~\cite{10.1145/3385416} also contains experimental results. However, their main focus is on hypergraphs and there are no comparisons against the algorithms mentioned above.

Aridhi \etal~\cite{aridhi2016distributed} gave a distributed $k$-core decomposition algorithm in large dynamic graphs. The authors used a graph partitioning approach to distribute the workload and pruning techniques to find nodes that are affected by the changes.
Wang \etal~\cite{DBLP:conf/icdcs/WangYJQXH17} gave a parallel algorithm that appears to significantly outperform the \Tool{TRAVERSAL} algorithm.
Jin \etal~\cite{DBLP:journals/tpds/JinWYHSX18} presented a parallel approach based on matching to update core numbers in fully dynamic networks. Specifically, the authors showed that if a batch of inserted/deleted edges forms a matching, 
then the core number update step can be performed in parallel. However, the type of the edges has to be the same (\ie{} only insertions, or only deletions) in each update.
Hua \etal~\cite{DBLP:journals/tpds/HuaSYJYCCC20} noted that previous algorithms become inefficient for high superior degree vertices, \ie{} vertices that have many neighbors that have a core number that is larger than its own core number. 
For example, the matching-based approach of Jin \etal\cite{DBLP:journals/tpds/JinWYHSX18} can only process one edge associated to a vertex in each iteration. Their new algorithm can handle multiple insertions/deletions per iteration. 

It would be interesting to evaluate the algorithm of Sun \etal~\cite{10.1145/3385416}  which maintains a $(4+\eps)$-approximate core number, on graphs to see how far from the exact core numbers these estimates are and how its running time compares to the above approaches. Note that an $(\alpha, \beta)$-decomposition actually gives a $(2\alpha + \eps)$ approximation and $\alpha$ has to be chosen to be slightly larger than 2 only to guarantee polylogarithmic updates. Thus, it would be interesting to also experiment with smaller values of $\alpha$.

 \ifEnableExtend
2010 ``K-shell decomposition for dynamic complex networks''~\cite{DBLP:conf/wiopt/MiorandiP10} \\
2014 ``Efficient Core Maintenance in Large Dynamic Graphs''~\cite{DBLP:journals/tkde/LiYM14} \\
2016 ``Incremental k-core decomposition: algorithms and evaluation''~\cite{DBLP:journals/vldb/SariyuceGJWC16} \\ 
2017 ``A Fast Order-Based Approach for Core Maintenance''~\cite{DBLP:conf/icde/ZhangYZQ17} \\
2017 ``Parallel Algorithm for Core Maintenance in Dynamic Graphs''~\cite{DBLP:conf/icdcs/WangYJQXH17} \\
2018 ``Core Maintenance in Dynamic Graphs:A Parallel Approach Based on Matching''~\cite{DBLP:journals/tpds/JinWYHSX18} \\
2020 ``Faster Parallel Core Maintenance Algorithms in Dynamic Graphs''~\cite{DBLP:journals/tpds/HuaSYJYCCC20}
 \fi{}

\subsection{Subgraph and Motif Counting}
Two graphs are isomorphic if there is a bijection between the vertex sets of
the graphs that preserves adjacency.
Given a graph pattern $H$, 
\emph{induced subgraph counting}, also called \emph{motif counting},
is concerned with the number of subgraphs of $G$ that are isomorphic to $H$,
divided by the number of automorphisms of $H$.
By contrast, non-adjacencies in $H$ 
need not be preserved in the \emph{non-induced} case.
For example, the non-induced count for so-called wedges (a path of length two)
in a triangle is three, whereas the induced count for the same pattern is zero.
A query for the induced or non-induced counting problem hence is answered by a
single numeric value.
Observe that induced and non-induced counts coincide if the pattern graph is a
clique.

In the \emph{detection} version of these problems, one is only
interested in whether the count is nonzero and upon a query, an
algorithm returns a Boolean value.
In other scenarios, the subgraphs also need to be \emph{enumerated}.
Here, a query is expected to return a sequence of subgraphs and
if running time is considered, the \emph{enumeration delay}, \ie{}
the time between the enumeration of a subgraph and its successor
in this sequence during answering a query, is of interest.

In the research that is currently available there is a subset of work that
focuses on the special case of counting wedges, triangles, as well as various
subgraph patterns on four vertices in dynamic networks.
Unless denoted otherwise, graphs and patterns are undirected in the following,
which is the more common setting.

\paragraph{Theory Results}
There is a conditional lower bound of $(\poly(n), m^{1/2-\epsilon},
m^{1-\epsilon})$ for any small constant $\epsilon >0$ even for the fundamental problem of detecting whether an
graph contains a triangle~\cite{henzinger2015unifying}.
The same lower bound also extends to various four-vertex subgraphs,
whereas there is a lower bound of $(\poly(n), m^{1-\epsilon},m^{2-\eps})$ for counting
$4$-cliques as well as \emph{induced} connected four-vertex subgraphs~\cite{subgraph-counting-sand}.
Interestingly, assuming the OMv-conjecture the problem of dynamically counting 4-cycles
remains hard also in the average case: Henzinger, Lincoln, and Saha~\cite{HenzingerLS22} defined a dynamic random graph model based on Erd{\H{o}}s-R{\'{e}}nyi random graphs, where the adversary can determine whether to update or query, but in case of an update a random pair of vertices is picked uniformly at random and the corresponding edge is ``flipped'', i.e., inserted if it does not exist and deleted if it does.
They gave a conditional lower bound of $(n^{3-\epsilon}, n^{1-\epsilon}, n^{2-\epsilon})$ for counting the number of 4-cycles in such a dynamic graph and a somewhat weaker polynomial bound for counting the number of triangles.

A fully dynamic algorithm with $\bigO(\sqrt{m})$ update time and constant query time
was recently given independently by Kara \etal~\cite{DBLP:conf/icdt/KaraNNOZ19,DBLP:journals/tods/KaraNNOZ20} for counting triangles.
Their algorithm is also able to enumerate triangles with $\bigO(1)$ delay.
Subsequently, Lu and Tao~\cite{DBLP:conf/icdt/LuT21} studied the trade-off between update time and approximation quality
and presented a new data structure for exact triangle counting whose complexity depends on the arboricity of the graph.
The results by Kara \ea{} were extended to general $k$-clique counting by Dhulipala \etal~\cite{DBLP:journals/corr/abs-2003-13585}.
Motivated by the fact that real-world graphs in certain applications often have small h-index $h$ (i.e., there are at most $h$ vertices of degree at least $h$),
Eppstein and Spiro~\cite{DBLP:journals/jgaa/EppsteinS12} showed that the undirected triangle count can be maintained in $\bigO(h)$ update time and constant query time.
Eppstein \ea{}~\cite{DBLP:journals/tcs/EppsteinGST12} later extended this
result to maintaining the counts of directed triangles in amortized $\bigO(h)$
update time and of undirected four-vertex subgraphs in amortized $\bigO(h^2)$ update time.
Note that $h$ can be as large as $\bigO(\sqrt{m})$, resulting in an amortized
time complexity of $\bigO(m)$ per update for four-vertex subgraphs in general.
Only very recently, Hanauer \ea{}~\cite{subgraph-counting-sand} showed
how to reduce this to amortized $\bigO(m^{2/3})$ time per update for all
four-vertex subgraphs except the $4$-clique.
The query time remains constant.

This is currently an active area of research.

\paragraph{Experimental Results on Triangle Counting}
Pavan \etal~\cite{DBLP:journals/pvldb/PavanTTW13} introduced neighborhood sampling to count and sample triangles in a one-pass streaming algorithm.
In neighborhood sampling, first a random edge in the stream is sampled and in subsequent steps, edges that share an endpoint with the already sampled edges are sampled. The algorithm outperformed their implementations of the previous best algorithms for the problem, namely the algorithms by Jowhari and Ghodsi~\cite{DBLP:conf/cocoon/JowhariG05} and by Buriol \etal~\cite{DBLP:conf/pods/BuriolFLMS06}.  
Note that the method does not appear to be able to handle edge deletions.
Bulteau \etal~\cite{DBLP:journals/algorithmica/BulteauFKP16} estimated the number of triangles in fully dynamic streamed graphs. Their method adapts 2-path sampling to work for dynamic graphs. The main idea of 2-path sampling is to sample a certain number of 2-paths and compute the ratio of 2-paths in the sample that are complete triangles. The total number of 2-paths in the graph is then multiplied with the ratio to obtain the total number of 2-paths in the graph. This approach fails, however, if one allows deletions. Thus, the contribution of the paper is a novel technique for sampling 2-paths. More precisely, the algorithm first streams the graph and sparsifies it. Afterwards, the sampling technique is applied on the sparsified graph. The core contribution of the authors is to show that the estimate obtained in the sparsified graph is similar to the number of triangles in the original graph.
For graphs with constant transitivity coefficient, \ie{} the ratio of 2-paths in $G$ contained in a triangle to all 2-paths in $G$, the authors achieve constant processing time per edge.
Makkar \etal~\cite{DBLP:conf/hipc/MakkarBG17} presented an exact and parallel approach using an inclusion-exclusion formulation for triangle counting in dynamic graphs. The algorithm is implemented in \Tool{cuSTINGER}~\cite{feng2015distinger} and runs on GPUs. The algorithm computes updates for batches of edge updates and also updates the number of triangles each vertex belongs to.
The \Tool{TRI{\`{E}}ST} algorithm~\cite{DBLP:journals/tkdd/StefaniERU17} estimates local and global triangles.  An input parameter of the algorithm is the amount of available memory. The algorithm maintains a sample of the edges using reservoir sampling and random pairing to exploit the available memory as much as possible. The algorithm reduces the average estimation error by up to \SI{90}{\percent} w.r.t.\ to the previous state-of-the-art.
Han and Sethu~\cite{DBLP:conf/asunam/HanS17} proposed a new sampling approach, called \Tool{edge-sample-and-discard}, which generates an unbiased estimate of the total number of triangles in a fully dynamic graph. The algorithm significantly reduces the estimation error compared to \Tool{TRI{\`{E}}ST}.
The \Tool{MASCOT} algorithm~\cite{DBLP:conf/kdd/LimK15,DBLP:journals/tkdd/LimJK18} focuses on local triangle counting, \ie{} counting the triangles adjacent to every node. In their work, the authors provide an unbiased estimation of the number of local triangles.

\paragraph{Experimental Results on Counting More Complex Patterns}
The neighborhood sampling method of Pavan \etal~\cite{DBLP:journals/pvldb/PavanTTW13} can also be used for more complex patterns, for example Pavan \etal also presented experiments for 4-cliques.
Shiller \etal~\cite{DBLP:conf/alcob/SchillerJHS15} presented the stream-based (insertions and deletions) algorithm \Tool{StreaM} for counting 4-vertex motifs in dynamic graphs. 
Ahmed \etal~\cite{DBLP:journals/pvldb/AhmedDWR17} presented a general purpose sampling framework for graph streams. The authors proposed a martingale formulation for non-induced subgraph count estimation and showed how to compute unbiased estimate of subgraph counts from a sample at any point during the stream. The estimates for triangle and wedge counting obtained are less than \SI{1}{\percent} away from the true number of triangles/wedges. The algorithm outperformed their own implementation of \Tool{TRI{\`{E}}ST} and \Tool{MASCOT}.
Mukherjee \etal~\cite{DBLP:journals/bmcsb/MukherjeeHBK18} gave an exact counting algorithm for a given set of motifs in dynamic networks. 
Their focus is on biological networks. The algorithm computes an initial embedding of each motif in the initial network.  Then for each motif its embeddings are stored in a list. This list is then dynamically updated while the graph evolves.

Dhulipala \etal~\cite{DBLP:journals/corr/abs-2003-13585} recently gave parallel batch-dynamic algorithms for $k$-clique counting and enumeration.
Their first algorithm is a batch-dynamic parallel algorithm for triangle counting that has amortized work $\bigO(\BatchSize \sqrt{\BatchSize + m})$ and $\bigO(\log^*(\BatchSize + m))$ depth with high probability for a batch of $\BatchSize$ edge insertions or deletions. The algorithm is based on degree thresholding which divides the vertices into vertices with low- and high-degree. Given the classification of the vertex, different update routines are used.
A multicore implementation of the triangle counting algorithm is given. Experiments indicate that the algorithms achieve \numrange{36.54}{74.73}-times parallel speedups on a machine with \num{72}~cores.
Lastly, the authors developed a simple batch-dynamic algorithm for enumerating $k$-cliques,
which has expected $\bigO(\BatchSize (m+\BatchSize)\alpha^{k-4})$ work and $\bigO(\log^{k-2} n)$ depth with high probability, for graphs with arboricity $\alpha$.

\ifEnableExtend
2013 ``Counting and Sampling Triangles from a Graph Stream''~\cite{DBLP:journals/pvldb/PavanTTW13} \\
2015 ``StreaM - {A} Stream-Based Algorithm for Counting Motifs in Dynamic Graphs''~\cite{DBLP:conf/alcob/SchillerJHS15} \\
2015 ``Mascot: Memory-efficient and accurate sampling for counting local triangles in graph streams''~\cite{DBLP:conf/kdd/LimK15,DBLP:journals/tkdd/LimJK18} \\
2016 ``Triangle Counting in Dynamic Graph Streams''~\cite{DBLP:journals/algorithmica/BulteauFKP16} \\
2017 ``Exact and Parallel Triangle Counting in Dynamic Graphs''~\cite{DBLP:conf/hipc/MakkarBG17}\\
2017 ``TRI{\`{E}}ST: Counting Local and Global Triangles in Fully Dynamic Streams with Fixed Memory Size''~\cite{DBLP:journals/tkdd/StefaniERU17} \\
2017 ``{Edge Sample and Discard: {A} New Algorithm for Counting Triangles in Large Dynamic Graphs}''~\cite{DBLP:conf/asunam/HanS17} \\
2017 ``On Sampling from Massive Graph Streams''~\cite{DBLP:journals/pvldb/AhmedDWR17} \\
2018 ``Counting motifs in dynamic networks''~\cite{DBLP:journals/bmcsb/MukherjeeHBK18} \\
2019 ``Sampling Methods for Counting Temporal Motifs''~\cite{DBLP:conf/wsdm/LiuBC19}
2020 ``Efficient Sampling Algorithms for Approximate Temporal Motif Counting''~\cite{DBLP:conf/cikm/00040JLT20}\\
\fi{}

To summarize for this problem the empirical work is far ahead of the
theoretical work and it would be interesting to better understand the
theoretical complexity of subgraph and motif counting.

\subsection{Eccentricity and Diameter}
The \emph{eccentricity} of a vertex is the greatest distance between the vertex and any other vertex in an undirected graph. 
Here, the distance between two vertices in a graph refers to the number of edges in a shortest path between those two vertices. For any vertex $v$ note that the depth of the BFS tree rooted at $v$ equals its eccentricity.
Based on this definition, the \emph{diameter} of a graph is defined as the maximum eccentricity over all vertices in the graph. Let $x$ and $y$ be two vertices whose distance $d(x,y)$ equals the graph diameter. Note that the eccentricity $ecc(v)$ of any vertex  $v$ is never larger and at most a factor 2 smaller than the diameter, as $2 \cdot ecc(v) \ge 2 \cdot \max(d(x,v), d(y,v)) \ge d(x,v) + d(v, y) \ge d(x,y)$.
The \emph{radius} is the minimum eccentricity of all vertices. Let $x$ be the vertex with minimum eccentricity,  and let $y$ be a vertex such that $d(x,y) = ecc(x)$. Note that the eccentricity of any vertex  $v$ is never smaller and at most a factor 2 larger than the radius, as $d(x,v) \le d(x,y) = ecc(x)$ and, thus,
$ecc(v) \le d(x,y) + ecc(x) \le 2 ecc(x)$.
Thus, through recomputation from scratch it is straightforward to compute a $2$-approximation for diameter and radius
in linear time.

\paragraph{Theory Results}
Anacona \etal~\cite{DBLP:conf/icalp/AnconaHRWW19} recently showed that under the strong exponential time hypothesis (SETH) there can be no $(2-\eps)$-approximate fully dynamic approximation algorithm for any of these problems with $\bigO(m^{1-\delta})$ update or query time for any $\delta > 0$.
There also exist non-trivial (and sub-$n^2$ time) fully dynamic algorithms for $(1.5+\epsilon)$ approximate diameter (and also for radius and eccentricities) \cite{DBLP:conf/focs/BrandNS19}. In that paper, the authors also construct a non-trivial algorithm for exact diameter.

We are not aware of any experimental study for fully dynamic diameter.
\ifEnableExtend
2019 ``Algorithms and Hardness for Diameter in Dynamic Graphs''~\cite{DBLP:conf/icalp/AnconaHRWW19}
\fi{}

\subsection{Independent Set and Vertex Cover}

Given a graph $G=(V,E)$, an \emph{independent set} is a set $S \subseteq V$ such that no vertices in $S$ are adjacent to one another. 
The \emph{maximum independent set problem} is to compute an independent set of  maximum cardinality, called a \emph{maximum independent set} (MIS).
The  \emph{minimum vertex cover} problem is 
equivalent to the maximum independent set problem:
$S$ is a minimum vertex cover $C$ in $G$ iff $V\setminus S$ is a maximum independent set $V\setminus C$ in $G$.
Thus, an algorithm that solves one of these problems can be used to solve the other. Note, however, that this does not hold for approximation algorithms: If $C'$ is an $\alpha$-approximation of a minimum vertex cover, then $V \setminus C'$ is not necessarily an $\alpha$-approximation of a maximum independent set.
Another related problem is the \emph{maximal} independent set problem. A set $S$ is a maximal independent set if it is an independent set such that for any vertex $v \in V \setminus S$,  $S \cup \{v\}$ is not independent.

\paragraph{Theory Results}
As computing the size of an MIS is NP-hard, all dynamic algorithms of independent set study the maximal independent set problem. Note, however, that unlike for matching a maximal independent set does not give an approximate solution for the MIS problem, as shown by a star graph.
In a sequence of papers~\cite{DBLP:conf/stoc/AssadiOSS18,DBLP:journals/corr/abs-1804-01823,DBLP:conf/soda/AssadiOSS19,DBLP:conf/focs/ChechikZ19,DBLP:conf/focs/BehnezhadDHSS19} the running time for 
the maximal independent set problem was reduced to $\bigO(\log^4 n)$ expected worst-case update time. All these algorithms actually maintain a maximal independent set. A query can either return the size of that set in constant time or output the whole set in time linear in its size.

\paragraph{Experimental Results}
While quite a large amount of engineering work has been devoted to the computation of independent sets/vertex covers in static graphs, the amount of engineering work for the dynamic independent set problem is very limited.
Zheng \etal~\cite{DBLP:conf/icde/ZhengWYC018} presented a heuristic fully dynamic algorithm and proposed a lazy search algorithm to improve the size of the maintained independent set.
A year later, Zheng \etal~\cite{DBLP:conf/icde/ZhengPCY19} improved the result such that the algorithm is less sensitive to the quality of the initial solution used for the evolving MIS. In their algorithm, the authors used two well known data reduction rules, degree one and degree two vertex reduction,  that are frequently used in the static case.   Moreover, the authors can handle batch updates. Bhore \etal~\cite{DBLP:conf/esa/Bhore0N20} focused on the special case of MIS for independent rectangles which is frequently used in map labelling applications. The authors presented a deterministic algorithm for maintaining a MIS of a dynamic set of uniform rectangles with amortized sub-logarithmic update time. Moreover, the authors evaluated their approach using extensive experiments.

\ifEnableExtend
2015 ``Deterministic Fully Dynamic Data Structures for Vertex Cover and Matching''~\cite{DBLP:conf/soda/BhattacharyaHI15} \\
2017 ``Fully Dyn. Appx. Maximum Matching and Minimum Vertex Cover in \emph{O}(log\({}^{\mbox{3}}\) \emph{n}) Worst Case Update Time''~\cite{DBLP:conf/soda/BhattacharyaHN17} \\
2017 ``Det. Fully Dynamic Approximate Vertex Cover and Fractional Matching in {O(1)} Amortized Update Time''~\cite{DBLP:conf/ipco/BhattacharyaCH17} \\
2018 ``Simple dynamic algorithms for Maximal Independent Set and other problems''~\cite{DBLP:journals/corr/abs-1804-01823} 
2018 ``Efficient Computation of a Near-Maximum Independent Set over Evolving Graphs'~\cite{DBLP:conf/icde/ZhengWYC018} \\
2018 ``Deterministic Fully Dynamic Data Structures for Vertex Cover and Matching''~\cite{DBLP:journals/siamcomp/BhattacharyaHI18} \\
2018 ``Fully dynamic maximal independent set with sublinear update time''~\cite{DBLP:conf/stoc/AssadiOSS18}\\
2019 ``Fully Dynamic Maximal Independent Set with Sublinear in n Update Time''~\cite{DBLP:conf/soda/AssadiOSS19} \\
2019 ``Computing a Near-Maximum Independent Set in Dynamic Graphs''~\cite{DBLP:conf/icde/ZhengPCY19} \\
2019 ``Fully Dynamic Maximal Independent Set in Expected Poly-Log Update Time''~\cite{DBLP:conf/focs/ChechikZ19} \\
2019 ``Fully Dynamic Maximal Independent Set with Polylogarithmic Update Time''~\cite{DBLP:conf/focs/BehnezhadDHSS19} \\
2020 ``An Algorithmic Study of Fully Dynamic Independent Sets for Map Labeling''~\cite{DBLP:conf/esa/Bhore0N20}\\
2020 ``Dynamic Approximate Maximum Independent Set of Intervals, Hypercubes and Hyperrectangles''~\cite{DBLP:conf/compgeom/Henzinger0W20} \\
\fi{}

\subsection{Shortest Paths}
One of the most studied problems on weighted dynamic networks is the
maintenance of shortest path information between pairs of vertices, where edge weights represent distances
and the length of a path is the sum of the weights of the edges on the path.
In the case of \emph{uniform edge weights} all weights are assumed to equal $1$.
Both the \emph{single-source} as well as the
in the \emph{all-pairs
shortest path problem} (\emph{APSP})
we are interested in
the shortest path between two arbitrary vertices $s$ and $t$ that are query parameters.
For the \emph{single-source shortest path problem} (\emph{SSSP}), the source vertex $s$ is
fixed beforehand and the dynamic graph algorithm is only required to answer
distance queries between $s$ and an (arbitrary) vertex $t$ which is specified
by the query operation.
In the \emph{$s$-$t$ shortest path problem} (\emph{STSP}) both $s$ and $t$ are fixed
beforehand and the data structure is only required to return the distance
between $s$ and $t$ as answer to a query.
There are two types of queries, namely a \emph{distance query}, which
returns the length of a shortest path and a
\emph{path reporting query}, which returns the shortest path itself.

The problem can be cast on both undirected or directed graphs:
In case of the latter, we are asking for a shortest path from $s$ to $t$ rather
than a shortest path between $s$ and $t$.
By replacing each edge in an undirected graph by a pair of directed,
anti-parallel edges, any algorithm for the directed case can also be used in
the undirected setting.
In some settings, the range of possible edge weights is restricted to positive
integers or to positive or non-negative real values.
Note that the length of a shortest path is undefined in the presence of
negative-length cycles.

\added{%
\paragraph{Journey Planning}
Algorithms for shortest path computations also play an important role in
graph-based journey planning approaches.
Here, the input is a timetable consisting of connections in, \eg{} a public
transportation system, which can then be modelled as a directed graph, and
queries ask, \eg{} for the earliest arrival time or the minimum number of
vehicle changes if one wants to travel from a given station to another and
depart not earlier than at a specified time.
Delays result in updates to the directed graph, which can be edge insertions,
edge deletions, and weight changes.
Different graph models have been proposed, such as the time-expanded model, the
time-dependent model, the reduced time-expanded
model~\cite{DBLP:journals/jea/PyrgaSWZ07}, and the dynamic timetable
model~\cite{DBLP:journals/jda/CioniniDDFGPZ17}.
In the time-dependent model, nodes represent stations and each edge represents
a connection between stations, where the corresponding edge weight is
time-dependent and encodes the timetable information.
In the other models, departure or arrival times are encoded within the nodes,
and a single update to the timetable may result in an entire set of
changes to the topology of the directed graph as well as edge weights.
}

\subsubsection{Theory Results}
Let $\delta > 0$ be any small constant. There is a conditional lower bound of
$(\poly(n,\allowbreak\log\MaxWeight),\allowbreak m^{1/2-\delta}, m^{1-\delta})$  based on the
OMv conjecture, even for $s$-$t$ shortest paths with uniform edge weights\cite{henzinger2015unifying}.
This lower bound applies also to any algorithm that gives a better than
$5/3$-approximation.
For planar graphs the product of query and update time is $\Omega(n^{1-\delta})$
based on the APSP conjecture~\cite{abboud2016popular}.
As even the partially dynamic versions have shown to be at least as hard as the
static all-pairs shortest paths
problem~\cite{DBLP:journals/algorithmica/RodittyZ11,abboud2016popular}, one
cannot hope for a \emph{combinatorial} fully dynamic all-pairs shortest paths
algorithm with $\bigO(n^{3-\delta})$ preprocessing time,
$\bigO(n^{2-\delta})$ amortized update time, and constant query time.

The state-of-the-art algorithms come close to this, even for the APSP problem.

For directed, weighted graphs,
Demetrescu and Italiano~\cite{DBLP:journals/jacm/DemetrescuI04} achieved an
\emph{amortized} update time of \added{$\bigO(n^2 \log^3 n)$}, which was later improved
\added{to $\bigO(n^2(\log n + \log^2((n+m)/n)))$} by Thorup~\cite{DBLP:conf/swat/Thorup04}.
Both of these algorithms actually allow vertex insertions and deletion, not just edge updates, but can only answer distance queries. 
In undirected graphs with uniform weight and
with preprocessing time $\bigO(n^2 \poly(\log n))$ a worst-case update time
of $\bigO(n^{1.9})$ can be achieved, however, with a $\bigO(n^{1.529})$ time per distance query and $\bigO(n^{1.9})$ per path reporting query\cite{DBLP:conf/soda/BergamaschiHGWW21}.

There is also a fully dynamic
$2^{\bigO(k^2)}$-approximation algorithm that takes time $\softO(\sqrt m n^{1/k})$ per update and $\bigO(k^2)$ per distance query for any positive integer $k$~\cite{DBLP:conf/approx/AbrahamCT14} and an $n^{o(1)}$-approximation with $n^{o(1)}$ update time~\cite{Forster2021}. These two algorithms only allow edge insertions and deletions. The second one can answer distance queries in constant time and path reporting queries in time linear in the number of edges on the approximate shortest path whose length it reports.

With respect to \emph{worst-case} update times, the currently fastest
algorithms are randomized with $\softO(n^{2+2/3})$ update
time per vertex update~\cite{DBLP:conf/soda/AbrahamCK17,DBLP:conf/soda/GutenbergW20b}.
Moreover, Probst Gutenberg and Wulff-Nilsen~\cite{DBLP:conf/soda/GutenbergW20b} presented a deterministic algorithm with $\softO(n^{2+5/7})$ update time,
thereby improving a \num{15} years old result by
Thorup~\cite{DBLP:conf/stoc/Thorup05}. 
Van den Brand and Nanongkai~\cite{DBLP:conf/focs/BrandN19} showed that
Monte Carlo-randomized $(1+\eps)$-approximation algorithms exist with
$\softO(n^{1.823}/\eps^2)$ worst-case update time for the fully dynamic
single-source shortest path problem and
$\softO(n^{2.045}/\eps^2)$ for
all-pairs shortest paths, in each case with positive real edge weights
and constant query time. However, both of these algorithm can only answer distance queries and not path-reporting queries.

Slightly faster exact and approximative algorithms exist in part for the
``special cases'' of uniform edge weights~\cite{%
DBLP:conf/cocoon/Sankowski05,%
DBLP:journals/algorithmica/RodittyZ11,%
DBLP:conf/soda/AbrahamCK17,%
DBLP:conf/soda/GutenbergW20b,%
DBLP:conf/focs/BrandNS19,%
DBLP:conf/focs/BrandN19%
} 
and/or \emph{undirected} graphs~\cite{%
DBLP:journals/siamcomp/RodittyZ12,%
DBLP:conf/focs/BrandN19%
} (every edge has a reverse edge of the same weight).
More details on shortest paths algorithms including fully dynamic algorithms are given in the
survey of Madkour et al.~\cite{madkour2017survey}.
%

\subsubsection{Experimental Results for Single-Source Shortest Paths}
The first experimental study for fully dynamic single-source shortest paths on
\emph{directed} graphs with positive real edge weights was conducted by Frigioni
\ea{}~\cite{DBLP:journals/jea/FrigioniINP98}, who evaluated Dijkstra's seminal
static algorithm~\cite{DBLP:journals/nm/Dijkstra59} against a fully dynamic
algorithm by Ramalingam and Reps~\cite{DBLP:journals/tcs/RamalingamR96}
(\Tool{RR}) as well as one by Frigioni
\ea{}~\cite{DBLP:conf/soda/FrigioniMN96} (\Tool{FMN}).
\Tool{RR} is based on Dijkstra's static algorithm and maintains a spanning
subgraph consisting of edges that belong to at least one shortest $s$-$t$ path
for some vertex $t$.
After an edge insertion, the spanning subgraph is updated starting from the
edge's head until all affected vertices have been processed.
In case of an edge deletion, the affected vertices are identified as a first
step, followed by an update of their distances.
The resulting worst-case update time is $\bigO(x_\delta + n_\delta \log
n_\delta) \subseteq \bigO(m + n \log n)$, where $n_\delta$ corresponds to the
number of vertices affected by the update, i.e., whose distance from $s$
changes and $x_\delta$ equals $n_\delta$ plus the number of edges incident to
an affected vertex.
Similarly, Frigioni \ea{}~\cite{DBLP:conf/soda/FrigioniMN96} analyzed the update
complexity of their algorithm \Tool{FMN} with respect to the change in the
solution and showed a worst-case running time of $\bigO(|U_\delta|\sqrt{m}\log
n)$, where $U_\delta$ is the set of vertices where either the distance from $s$
must be updated or their parent in the shortest paths tree.
The algorithm assigns each edge $(u, v)$ a forward (backward) level, which
corresponds to the difference between the (sum of) $v$'s ($u$'s) distance from
$s$ and the edge weight, as well as an owner, which is either $u$ or $v$, and
used to bound the running time.
Incident outgoing and incoming edges of a vertex that it does not own are kept
in a priority queue each, with the priority corresponding to the edge's level.
In case of a distance update at a vertex, only those edges are scanned that
are either owned by the vertex or have a priority that indicates a shorter
path.
Edge insertion and deletion routines are based on Dijkstra's algorithm and
handled similar as in \Tool{RR}, but using level and ownership information.
The experiments were run on three types of input instances: randomly generated
ones, instances crafted specifically for the tested algorithms, and random
updates on autonomous systems networks.
The static Dijkstra algorithm is made dynamic in that it is re-run from scratch
each time its shortest paths tree is affected by an update.
The evaluation showed that the dynamic algorithms can speed up the update time
by \SI{95}{\percent} over the static algorithm.
Furthermore, \Tool{RR} turned out to be faster in practice than \Tool{FMN}
except on autonomous systems instances, where the spanning subgraph was large
due to many alternative shortest paths.

In a follow-up work, Demetrescu
\ea{}~\cite{DBLP:conf/wae/DemetrescuFMN00,Demetrescu01fullydynamic} extended
this study to dynamic graphs with arbitrary edge weight, allowing in particular
also for negative weights.
In addition to the above mentioned algorithm by Ramalingam and
Reps~\cite{DBLP:journals/tcs/RamalingamR96} in a slightly
lighter version (\Tool{RRL}) and the one by Frigioni
\ea{}~\cite{DBLP:conf/soda/FrigioniMN96} (\Tool{FMN}), their study also
includes a simplified variant of the latter which waives edge ownership
(\Tool{DFMN}), as well as a rather straightforward dynamic algorithm
(\Tool{DF}) that in case of a weight increase on an edge $(u, v)$ first marks
all vertices in the shortest paths subtree rooted at $v$ and then finds for
each of these vertices an alternative path from $s$ using only unmarked
vertices.
The new weight of these vertices can be at most this distance or
the old distance plus the amount of weight increase on $(u, v)$.
Therefore, the minimum is taken as a distance estimate for the second step,
where the procedure is as in Dijkstra's algorithm.
In case of a weight decrease on an edge $(u, v)$ the first step is omitted.
As Dijkstra's algorithm is employed as a subroutine, the worst-case running
time of \Tool{DF} for a weight change is $\bigO(m + n \log n)$.
For updates, all algorithms use a technique introduced by Edmonds and
Karp~\cite{DBLP:journals/jacm/EdmondsK72} to transform the weight $w(u, v)$ of
each edge $(u, v)$ to a non-negative one by replacing it with the reduced weight
$w(u, v) - (d(v) - d(u))$, where $d(\cdot)$ denotes the distance from $s$.
This preserves shortest paths and allows Dijkstra's algorithm to be used during
the update process.
The authors compared these dynamic algorithms to re-running the static algorithm
by Bellman and Ford on each update from scratch on various randomly generated
dynamic instances with mixed incremental and decremental updates on the edge
weights, always avoiding negative-length cycles.
Their study showed that \Tool{DF} is the fastest in practice
on most instances, however, in certain circumstances \Tool{RR} and \Tool{DFMN}
are faster, whereas \Tool{FMN} turned out to be too slow in practice due to its
complicated data structures.
The authors observed a runtime dependency on the interval size of the edge
weights; \Tool{RR} was the fastest if this interval was small, except for very
sparse graphs.
\Tool{DFMN} on the other hand was shown to perform better than \Tool{DF} in
presence of zero-length cycles, whereas \Tool{RR} is incapable of handling such
instances.
It is interesting to note here that the differences in running time are only
due to the updates that increase distances, as all three candidates used
the same routine for operations that decrease distances.
The static algorithm was slower than the dynamic algorithms by several orders
of magnitude.

Buriol \ea{}~\cite{DBLP:journals/informs/BuriolRT08} presented a technique that
reduces the number of elements that need to be processed in a heap after an
update for various dynamic shortest paths algorithms by excluding vertices
whose distance changes by exactly the same amount as the weight change and
handling them separately.
They showed how this improvement can be incorporated into
\Tool{RR}~\cite{DBLP:journals/tcs/RamalingamR96},
a variant similar to \Tool{RRL}~\cite{DBLP:journals/tcs/RamalingamR96},
the algorithm by King and Thorup~\cite{DBLP:conf/cocoon/KingT01} (\Tool{KT}),
and \Tool{DF}~\cite{Demetrescu01fullydynamic}
and achieves speedups of up to \num{1.79} for random weight changes and up to
\num{5.11} for unit weight changes.
Narváez \ea{}~\cite{DBLP:journals/ton/NarvaezST00}
proposed a framework to dynamize static shortest path algorithms such as
Dijkstra's or Bellman-Ford~\cite{bellman1958routing}.
In a follow-up work~\cite{DBLP:journals/ton/NarvaezST01},
they developed a new algorithm that fits in this framework
and is based on the linear programming formulation of shortest paths and its
dual, which yields the problem in a so-called \emph{ball-and-string model}.
The authors experimentally showed that their algorithm needs fewer comparisons
per vertex when processing an update than the algorithms from their earlier
work, as it can reuse intact substructures of the old shortest path tree.

Misra and Oommen~\cite{DBLP:journals/tsmc/MisraO05} presented algorithms for
single-source shortest paths that are based on learning automata and designed to
find ``statistical'' shortest paths in a stochastic graph with stochastically
changing edge weights.
The algorithms are extensions of
\Tool{RR}~\cite{DBLP:journals/tcs/RamalingamR96} and
\Tool{FMN}~\cite{DBLP:conf/soda/FrigioniMN96}
and shown to be superior to the original versions of \Tool{RR} and \Tool{FMN}
by several orders of magnitude once they have converged.
Chan and Yang~\cite{DBLP:journals/tc/ChanY09} studied the problem of dynamically
updating a single-source shortest path tree under multiple concurrent edge weight
updates.
They amended the algorithm by Narváez \ea{}~\cite{DBLP:journals/ton/NarvaezST01} (\Tool{MBS}),
for which they showed that it may misbehave in certain circumstances and suggested
two further algorithms:
\Tool{MFP} is an optimized version of an algorithm by Ramalingam and
Reps~\cite{DBLP:journals/jal/RamalingamR96} (\Tool{DynamicSWSF-FP}), which can
handle multiple updates at once.
The second algorithm is a generalization of the dynamic Dijkstra algorithm
proposed by Narváez \ea{}~\cite{DBLP:journals/ton/NarvaezST00}.
In a detailed evaluation, they showed that an algorithm obtained by combining
the incremental phase of \Tool{MBS} and the decremental phase of their
dynamization of Dijkstra's algorithm performed best on road networks, whereas
the dynamized Dijkstra's algorithm was best on random networks.

An extensive experimental study on single-source shortest path algorithms was
conducted by Bauer and Wagner~\cite{DBLP:conf/wea/BauerW09}.
They suggested several tuned variants of
\Tool{DynamicSWSF-FP}~\cite{DBLP:journals/jal/RamalingamR96} and evaluated them
against \Tool{FMN}~\cite{DBLP:conf/soda/FrigioniMN96},
different algorithms from the framework by Narváez \ea{}~\cite{DBLP:journals/ton/NarvaezST00},
as well as \Tool{RR}~\cite{DBLP:journals/tcs/RamalingamR96} on a diverse set of instances.
The algorithms from the Narváez framework showed similar performance
in case of single-edge updates and were the fastest on road networks
and generated grid-like graphs.
By contrast, the tuned variants of \Tool{DynamicSWSF-FP} behaved less
consistent.
\Tool{RR} was superior on Internet networks, whereas \Tool{FMN} was the
slowest, especially on sparse instances.
Interestingly, the authors showed that for batch updates with a set of randomly
chosen edges, the algorithms behave similar as for single-edge updates, as
there was almost no interference.
The picture changed slightly for simulated node failures and strongly for
simulated traffic jams.
\Tool{RR} and a tuned variant of \Tool{DynamicSWSF-FP} showed the best
performance for simulated node failures, and two tuned variants of
\Tool{DynamicSWSF-FP} dominated in case of simulated traffic jams.
Notably, the algorithms from the Narváez framework were faster here if
instead of in batches, the updates were processed one-by-one.

In follow-up works, D'Andrea
\ea{}~\cite{
DBLP:journals/jea/DAndreaDF0P15}
evaluated several batch-dynamic algorithms for single-source shortest paths,
where the batches are homogeneous, i.e., all updates are either incremental or
decremental.
Their study contains \Tool{RR}~\cite{DBLP:journals/tcs/RamalingamR96},
a tuned variant of \Tool{DynamicSWSF-FP}~\cite{DBLP:journals/jal/RamalingamR96}
described by Bauer and Wagner~\cite{DBLP:conf/wea/BauerW09} (\Tool{TSWSF}),
as well as
a new algorithm \Tool{DDFLP}, which is designed specifically to handle
homogeneous batches and uses similar techniques as
\Tool{FMN}~\cite{DBLP:conf/soda/FrigioniMN96}.
The instance set comprised road and Internet networks as well as randomly
generated graphs according to the Erdős-Rényi model (uniform degree
distribution) and the Barabási-Albert model (power-law degree distribution).
Batch updates were obtained from simulated node failure and recovery,
simulated traffic jam and recovery, as well as randomly selected edges
for which the weights were either increased or decreased randomly.
The evaluation confirmed the results by Bauer and
Wagner~\cite{DBLP:conf/wea/BauerW09} and showed that \Tool{DDFLP} and
\Tool{TSWSF} are best in case of update scenarios like node failures or traffic
jams and otherwise \Tool{TSWSF} and \Tool{RR}, where \Tool{RR} is preferable
to \Tool{TSWSF} if the interference among the updates is low and vice versa.
\Tool{DDFLP} generally benefited from dense instances.

Singh and Khare~\cite{DBLP:journals/iajit/SinghK19} presented the first
batch-dynamic parallel algorithm for single-source shortest paths for GPUs and
showed in experiments that it outperforms the (sequential) tuned
\Tool{DynamicSWSF-FP} algorithm~\cite{DBLP:conf/wea/BauerW09} by a factor of up
to \num{20} if the distances of up to \SI{10}{\percent} of the nodes are
affected.

\subsubsection{Experimental Results for All-Pairs Shortest Paths}
The first fully dynamic algorithm for all-pairs shortest paths in graphs with
positive integer weights less than a constant $\MaxWeight$
was presented by
King~\cite{DBLP:conf/focs/King99} \added{(\Tool{Ki})} and
\added{later also evaluated experimentally.
It has} an amortized update time of
$\bigO(n^{2.5}\sqrt{\MaxWeight \log n})$.
For each vertex $v$, \added{\Tool{Ki}} maintains two shortest paths trees up to a distance $d$:
one outbound with $v$ as source and one inbound with $v$ as target.
A so-called stitching algorithm is used to stitch together longer paths from
shortest paths of distance at most $d$.
To achieve the above mentioned running time, $d$ is set to $\sqrt{n \MaxWeight \log n}$.
The space requirement is $\bigO(n^3)$ originally, but can be reduced to
$\softO(n^2\sqrt{n\MaxWeight})$~\cite{DBLP:conf/cocoon/KingT01}.

For non-negative, real-valued edge weights, Demetrescu and
Italiano~\cite{DBLP:journals/jacm/DemetrescuI04} proposed \added{the above-mentioned}
algorithm \added{(\Tool{DI})}
with an amortized update time of \added{$\softO(n^2)$}.
The algorithm uses the concept of \emph{locally shortest paths}, which are
paths such that each proper subpath is a shortest path, but not necessarily the
entire path, and \emph{historical shortest paths}, which are paths that have
once been shortest paths and whose edges have not received any weight updates
since then.
The combination of both yields so-called locally historical paths, which are
maintained by the algorithm.
To keep their number small, the original sequence of updates is transformed
into an equivalent, but slightly longer \emph{smoothed sequence}.
In case of a weight update, the algorithm discards all maintained paths
containing the updated edge and then computes new locally historical paths
using a routine similar to Dijkstra's algorithm.
Both \added{\Tool{Ki} and \Tool{DI}} have constant query time and were evaluated experimentally
in a study by Demetrescu and Italiano~\cite{DBLP:journals/talg/DemetrescuI06}
against \Tool{RRL}~\cite{DBLP:conf/wae/DemetrescuFMN00}
on random instances, graphs with a single bottleneck edge, which serves as a bridge
between two equally-sized complete bipartite graphs and only its weight is subject
to updates, as well as real-world instances obtained from the US road networks
and autonomous systems networks.
Apart from \Tool{RRL}, the study also comprises the Dijkstra's static algorithm.
Both these algorithms are designed for single-source shortest paths and were
hence run once per vertex.
All algorithms were implemented with small deviations from their respective
theoretical description to speed them up in practice.
The study showed that \Tool{RRL} and the algorithm based on locally historical
paths (\Tool{LHP}) can outperform the static algorithm by a factor of up to
\num{10000}, whereas the algorithm by King only achieved a speedup factor
of around \num{10}.
\Tool{RRL} turned out to be especially fast if the solution changes only
slightly, but by contrast exhibited the worst performance on the bottleneck
instances unless the graphs were sparse.
In comparison, \Tool{LHP} was slightly slower on sparse instances, but
could beat \Tool{RRL} as the density increased.
The authors also point out differences in performance that depend mainly on the
memory architecture of the machines used for benchmarking, where \Tool{RRL}
could better cope with small caches or memory bandwidth due to its reduced
space requirements and better locality in the memory access pattern, whereas
\Tool{LHP} benefited from larger caches and more bandwidth.

To speed up shortest paths computations experimentally, Wagner
\ea{}~\cite{DBLP:journals/jea/WagnerWZ05} introduced a concept for pruning the
search space by \emph{geometric containers}.
Here, each edge $(u, v)$ is associated with a set of vertices called container,
which is a superset of all vertices $w$ whose shortest $u$-$w$ path
starts with $(u, v)$.
The authors assume that each vertex is mapped to a point in two-dimensional
Euclidean space and based on this, suggest different types of geometric objects
as containers, such as disks, ellipses, sectors or boxes.
All types of container only require constant additional space per edge.
The experimental evaluation on static instances obtained from road and railway
networks showed that using the bounding box as container reduces the query time
the most in comparison to running the Dijkstra algorithm without pruning, as
the search space could be reduced to \SIrange{5}{10}{\percent}.
This could be preserved for dynamic instances obtained from railway networks if
containers were grown and shrunk in response to an update, with a speedup
factor of \numrange{2}{3} over a recomputation of the containers from scratch.
For bidirectional search, reverse containers need to be maintained
additionally, which about doubled the absolute update time.

Delling and Wagner~\cite{DBLP:conf/wea/DellingW07} adapted the static
\Tool{ALT} algorithm~\cite{DBLP:conf/soda/GoldbergH05} to the dynamic setting.
\Tool{ALT} is a variant of bidirectional $A^*$ search that uses a small subset
of vertices called \emph{landmarks}, for which distances from and to all other
vertices are precomputed, and the triangle inequality to direct the search
for a shortest path towards the target more efficiently.
The authors distinguish between an eager and a lazy dynamic version of
\Tool{ALT}, where the eager one updates all shortest path trees of the
landmarks immediately after an update.
The lazy variant instead keeps the preprocessed information as long as it still
guarantees correctness, which holds as long as the weight of an edge is at
least its initial weight, however at the expense of a potentially larger search
space.
The choice of landmarks remains fixed.
The experimental study on large road networks showed that queries in the lazy
version are almost as fast as in the eager version for short distances or if no
edges representing motorways are affected, but slower by several factors for
longer distances, larger changes to the weight of motorway edges, or after
many updates.

Schultes and Sanders~\cite{DBLP:conf/wea/SchultesS07} combined and generalized
different techniques that have been successfully used in the static setting,
such as separators, highway hierarchies, and transit node routing in a
multi-level approach termed \emph{highway-node routing}:
For the set of vertices $V_i$ on each level $i$, $V_i \subseteq V_{i-1}$, and the
overlay graph $G_i$ is defined on $V_i$ with an edge $(s, t) \in V_i \times
V_i$ iff there is a shortest $s$-$t$ path in $G_{i-1}$ that contains no
vertices in $V_{i}$ except for $s$ and $t$.
Queries are carried out by a modified Dijkstra search on this graph hierarchy.
The authors extended this approach also to the dynamic setting and consider
two scenarios: a server scenario, where in case of edge weight changes
the sets of highway nodes $V_i$ are kept and the graphs $G_i$ are updated,
and a mobile scenario, where only those vertices that are potentially affected
are determined and the query routine needs to be aware of possibly outdated
information during a search.
In an experimental evaluation on a very large road network with dynamically
changing travel times as weights it is shown that the dynamic highway-node
routing outperformed recomputation from scratch as well as dynamic \Tool{ALT}
search with \num{16} landmarks clearly with respect to preprocessing,
update, and query time as well as space overhead.

For real-time shortest path computations on networks with fixed topology, but
varying metric, Delling \ea{}~\cite{DBLP:conf/wea/DellingGPW11} suggested a
three-stage approach:
In the first, preprocessing step, a metric-independent, moderate amount of
auxiliary data is obtained from the network's topology.
It is followed by a customization step, which is run for each metric
and produces few additional data.
Whereas the first phase is run only once and can therefore use more computation
time, the second phase must complete within seconds in real-life scenarios.
Shortest path queries form the third phase and must be fast enough for actual
applications.
For the first, metric-independent stage, the authors describe an approach based
on graph partitioning, where the number of \emph{boundary edges}, i.e., edges
between different partitions, is to be minimized.
For the second stage, they compute an overlay network consisting of shortest
paths between all pairs of \emph{boundary nodes}, i.e. nodes that are incident to at
least one boundary edge.
An $s$-$t$ query is then answered by running a bidirectional Dijkstra algorithm
on the graph obtained by combining the overlay graph with the subgraphs induced
by the partitions containing $s$ and $t$, respectively.
The authors also considered various options for speedups, such as a
sparsification of the overlay network, incorporating goal-directed search
techniques, and multiple levels of overlays.
An experimental evaluation on road networks with travel distances and travel
times as metrics showed that their approach allows for real-time queries and
needs only few seconds for the metric-dependent customization phase.

\emph{Arc flags} belong in the category of goal-directed techniques to speed up
shortest path computations and have been successfully used in the static
setting~\cite{DBLP:journals/jea/BauerDSSSW10}.
To this end, the set of vertices is partitioned into a number of regions.
Each edge receives a label consisting of a flag for each region, which tells
whether there is a shortest path starting with this edge and ending in the
region.
The technique is related to geometric containers and uses the arc flags to
prune a (bidirectional) Dijkstra search.
Berettini \ea{}~\cite{DBLP:conf/atmos/BerrettiniDD09} were the first to
consider arc flags in a dynamic setting, however only for the case of weight
increases.
Their main idea is to maintain a threshold for each edge and region
that gives the increase in weight required for the edge to lie on a
shortest path.
On a weight increase, the thresholds are updated and used to determine when to
change an arc flag.
Although this potentially reduces the quality of the arc flags with each
update, the experimental evaluation showed that the increase in query time is
very small as long as the update sequence is short.
With respect to the update time, a significant speedup could be achieved over
recomputing arc flags from scratch.

To refresh arc flags more exactly and in a fully dynamical setting, D'Angelo
\ea{}~\cite{DBLP:journals/networks/DAngeloDF14} introduced a data structure
called \emph{road signs}.
Road signs complement arc flags and store for each edge $e$ and region $R$ the
set of boundary nodes contained in any shortest path starting with $e$ and
ending in $R$.
In case of a weight increase, the algorithm first identifies all affected nodes
whose shortest path to a boundary node changed and then updates all road signs
for all outgoing edges of an affected node.
In case of a weight decrease on edge $(u, v)$, the authors observed that all
shortest paths containing $(u, v)$ remain unchanged.
However, shortest paths starting with other outgoing or incoming edges of $u$
might require updates, as well as other paths containing an incoming edge of
$u$.
In an experimental study on road networks, the authors compared their algorithm
against one that recomputes arc flags from scratch as well as the algorithm by
Berettini \ea{}~\cite{DBLP:conf/atmos/BerrettiniDD09} (\Tool{BDD}).
To mimic traffic jams and similar occurrences, the weight of a randomly chosen
edge increases and then decreases by the same amount, however not necessarily
in subsequent updates.
The evaluation showed that updating both road signs and arc flags is by several
factors faster than recomputing arc flags from scratch.
On instances with weight increases only, the authors showed that their new
algorithm outperforms \Tool{BDD} distinctly both for updates and queries.

A further speedup technique for shortest path queries are \emph{2-hop cover
labelings}, where the label $L(v)$ of each node $v$ is a carefully chosen set
of nodes $U_v$ along with the distance between $v$ and $u$ for each $u \in U$.
For each pair of vertices $s$ and $t$, the shortest $s$-$t$ path can be
obtained by intersecting $U_s$ and $U_t$ and taking the minimum over all
combinations of $s$-$x$ and $x$-$t$ paths for all nodes $x \in U_s \cap U_t$.
In the static setting, a 2-hop cover labeling can be computed based on a
breadth-first search that is run once for every vertex (``naive landmark
labeling'').
Akiba \ea{}~\cite{DBLP:conf/www/AkibaIY14} introduced \emph{pruned landmark
labeling} (PLL), which constitutes a more refined approach and uses pruned
breadth-first searches instead.
The authors developed an incremental algorithm for PLL, which was complemented
by D'Angelo \ea{}~\cite{
DBLP:journals/jea/DAngeloDF19} to a fully dynamic algorithm.
The experimental evaluation showed that the algorithm achieves speedups of
several orders of magnitude over a recomputation from scratch, while at the
same time preserving the quality of the labeling, which makes this speedup
technique suitable for practical use in dynamic scenarios.

Hayashi \ea{}~\cite{DBLP:conf/cikm/0002AK16} proposed a method to support
shortest paths queries on unweighted networks with billions of edges by
combining a bidirectional breadth-first search, which is optimized for the
structure of small-world networks, with landmarks.
To this end, the authors choose high-degree vertices and store shortest path
trees as well as those of a subset of their neighbors in a so-called
``bit-parallel'' form.
This increases the number of landmarks, which in turn generally speeds up the
search and in particular for high-degree vertices, and at the same time keeps
the memory requirements comparatively small.
After an edge insertion or deletion, the bit-parallel shortest paths trees are
updated accordingly.
The experimental evaluation on twelve real-world instances having between
\num{1.5} million and \num{3.7} billion edges showed that the new algorithm was
able to process queries on average in less than \SI{8}{\milli\second} and even
considerably less on many instances.
The average edge insertion and deletion times were less than
\SI{1.3}{\milli\second} and \SI{8.1}{\second}, respectively, after an
initialization time of less than \SI{1}{\hour}.
The incremental algorithm by Akiba \ea{}~\cite{DBLP:conf/www/AkibaIY14}, which
was included in the study, was distinctly faster on queries, but on some
instances several factors slower on insertions.
However, it failed to complete the preprocessing step within \SI{10}{\hour} or
required more than \SI{128}{\giga\byte} of memory on half of all instances.


\subsubsection{Experimental Results for Journey Planning}
Cionini \etal{}~\cite{DBLP:journals/jda/CioniniDDFGPZ17} engineered
different update and query algorithms for the reduced time-expanded
and the dynamic timetable model and evaluated them on
public transportation timetables of different sizes.
They showed that update times are negligible in practice and that their new
heuristic query algorithms combined with \Tool{ALT} outperform other approaches
and is at least competitive with array-based models.
For the reduced time-expanded model, the algorithm outperforms one of the
fastest array-based algorithms, however at the expense of a larger memory
footprint in comparison to the dynamic timetable model.
The query time is in the order of milliseconds even for large timetables
with several millions of connections.
Giannakopoulou \etal{}~\cite{DBLP:journals/algorithms/GiannakopoulouP19}
extended the dynamic timetable model to the multi-modal journey planning problem,
which can deal with more than one transport mode.
In an experimental study, they showed that their algorithms answer
different kinds of queries effectively and are competitive to
state-of-the-art approaches.

\ifEnableExtend
1996 ``On the computational complexity of dynamic graph problems''~\cite{DBLP:journals/tcs/RamalingamR96} \\ 
1996 ``An Incremental Algorithm for a Generalization of the Shortest-Path Problem''~\cite{DBLP:journals/jal/RamalingamR96} \\ 
1996 ``Fully Dynamic Output Bounded Single Source Shortest Path Problem''~\cite{DBLP:conf/soda/FrigioniMN96}\\ 
1998 ``Fully Dynamic Shortest Paths and Negative Cycles Detection on Digraphs with Arbitrary Arc Weights''~\cite{DBLP:conf/esa/FrigioniMN98} \\ 
1998 ``Experimental Analysis of Dynamic Algorithms for the Single-Source Shortest-Path Problem''~\cite{DBLP:journals/jea/FrigioniINP98}\\

2000 ``Maintaining Shortest Paths in Digraphs with Arbitrary Arc Weights: An Experimental Study'~\cite{DBLP:conf/wae/DemetrescuFMN00}\\
2001 ``Temporal shortest paths: Parallel computing implementations''~\cite{DBLP:journals/pc/TremblayF01}\\ 
1999 ``Fully Dynamic Algorithms for Maintaining All-Pairs Shortest Paths and Transitive Closure in Digraphs''~\cite{DBLP:conf/focs/King99} \\ 
2001 ``Fully Dynamic Algorithms for Path Problems''~\cite{Demetrescu01fullydynamic} 

2000 ``New dynamic algorithms for shortest path tree computation''~\cite{DBLP:journals/ton/NarvaezST00}\\ 
2001 ``New dynamic {SPT} algorithm based on a ball-and-string model''~\cite{DBLP:journals/ton/NarvaezST01} \\

2004 ``A  new  approach  to  dynamic  all  pairs  shortest  paths''~\cite{DBLP:journals/jacm/DemetrescuI04} \\ 
2004 ``Fully-Dynamic All-Pairs Shortest Paths: Faster and Allowing Negative Cycles''~\cite{DBLP:conf/swat/Thorup04}\\
2005 ``Geometric containers for efficient shortest-path computation''~\cite{DBLP:journals/jea/WagnerWZ05}\\
2005 ``Dynamic algorithms for the shortest path routing problem: learning automata-based solutions''~\cite{DBLP:journals/tsmc/MisraO05} \\
2006 ``Fully dynamic all pairs shortest paths with real edge weights''~\cite{DBLP:journals/jcss/DemetrescuI06}\\ 
2006 ``Experimental analysis of dynamic all pairs shortest path algorithms''~\cite{DBLP:journals/talg/DemetrescuI06} \\

2007 ``Dynamic Highway-Node Routing''~\cite{DBLP:conf/wea/SchultesS07} \\
2007 ``Landmark-Based Routing in Dynamic Graphs''~\cite{DBLP:conf/wea/DellingW07} \\

2008 ``Speeding Up Dynamic Shortest-Path Algorithms''~\cite{DBLP:journals/informs/BuriolRT08} \\ 
2008 ``Finding time-dependent shortest paths over large graphs''~\cite{DBLP:conf/edbt/DingYQ08} \\ 
2008 ``Bidirectional Core-Based Routing in Dynamic Time-Dependent Road Networks''~\cite{DBLP:conf/isaac/DellingN08} \\ 
2009 ``Engineering Route Planning Algorithms''~\cite{DBLP:conf/dfg/DellingSSW09} \\ 
2009 ``Shortest Path Tree Computation in Dynamic Graphs''~\cite{DBLP:journals/tc/ChanY09} \\ 
2009 ``Batch dynamic single-source shortest-path algorithms: An experimental study''~\cite{DBLP:conf/wea/BauerW09} \\

2009 ``Arc-Flags in Dynamic Graphs''~\cite{DBLP:conf/atmos/BerrettiniDD09} \\

2011 ``Customizable Route Planning''~\cite{DBLP:conf/wea/DellingGPW11}\\
2011 ``Dynamic Arc-Flags in Road Networks''~\cite{DBLP:conf/wea/DAngeloFV11} \\ 
2012 ``Fully Dynamic Maintenance of Arc-Flags in Road Networks''~\cite{DBLP:conf/wea/DAngeloDFV12} \\ 

2013 ``Query Processing on Temporally Evolving Social Data''~\cite{DBLP:phd/basesearch/Huo13a} \\ 
2013 ``Dynamically maintaining shortest path trees under batches of updates''~\cite{DBLP:conf/sirocco/DAndreaDFLP13} \\ 
2014 ``Experimental evaluation of dynamic shortest path tree algorithms on homogeneous batches''~\cite{DBLP:conf/wea/DAndreaDFLP14} \\

2014 ``Efficient temporal shortest path queries on evolving social graphs''~\cite{DBLP:conf/ssdbm/HuoT14}\\ 
2014 ``Engineering shortest-path algorithms for dynamic networks''~\cite{DBLP:conf/ictcs/DEmidioF14} \\ 
2014 ``Dyn.\ and historical shortest-path distance queries on large evolving networks by pruned landmark labeling''~\cite{DBLP:conf/www/AkibaIY14} \\ 
2014 ``Fully dynamic update of arc-flags''~\cite{DBLP:journals/networks/DAngeloDF14} \\

2015 ``Dynamic Maintenance of a Shortest-Path Tree on Homogeneous Batches of Updates: New Algorithms and Experiments''~\cite{DBLP:journals/jea/DAndreaDF0P15}\\ 

2016 ``Distance queries in large-scale fully dynamic complex networks''~\cite{DBLP:conf/iwoca/DAngeloDF16} \\ 
2016 ``Shortest Paths on Evolving Graphs''~\cite{DBLP:conf/csonet/ZouZWLSZL16}\\ 
2016 ``Fully Dynamic Shortest-Path Distance Query Acceleration on Massive Networks''~\cite{DBLP:conf/cikm/0002AK16} \\ %

2017 ``Engineering graph-based models for dynamic timetable information systems''~\cite{DBLP:journals/jda/CioniniDDFGPZ17} \\ 

2018 ``Single-Source Shortest Path Tree for Big Dynamic Graphs''~\cite{DBLP:conf/bigdataconf/Riazi0DBN18} \Comment{distributed}\\ 
2019 ``Fully Dynamic 2-Hop Cover Labeling''~\cite{DBLP:journals/jea/DAngeloDF19} \\ 

2019 ``Parallel batch dynamic single source shortest path algorithm and its implementation on {GPU} based machine''~\cite{DBLP:journals/iajit/SinghK19} \\

2019 ``Multimodal Dynamic Journey-Planning''~\cite{DBLP:journals/algorithms/GiannakopoulouP19}\\ 
\fi{}
\subsection{Maximum Flows and Minimum Cuts}
An instance of the maximum flow/minimum cut problem consists of an
edge-weighted directed graph $G = (V, E, \Weight)$ along with two distinguished
vertices $s$ and $t$.
The edge weights $\Weight$ are positive and commonly referred to as
\emph{capacities}.
An ($s$-$t$) \emph{flow} $f$ is a non-negative weight function on the edges
such that $f(e) \leq \Weight(e)$ for all $e \in E$ (capacity constraints) and except for
$s$ and $t$, the total flow on the incoming edges of each vertex must equal the
total flow on the outgoing edges (conservation constraints).
The \emph{excess} of a vertex $v$ is the total flow on its incoming edges minus
that on its outgoing edges, which must be zero for all vertices except $s$ and
$t$.
The value of a flow $f$ then is the excess of $t$.
The task is to find a flow of maximum value.

An ($s$-$t$) \emph{cut} is a
partition $(S, T)$ of the set of vertices where $s \in S$ and $t \in T$.
Its value is the sum of the capacities of all edges $(u, v)$ such that $u \in
S$ and $v \in T$.
The well-known max-flow min-cut theorem states that the maximum value of a flow
equals the minimum value of a cut.

The \emph{global minimum cut} problem for an undirected edge-weighted graph
asks us to divide its set of nodes into two blocks while minimizing the weight
sum of the cut edges.

\paragraph{Theory Results}
In the dynamic setting,  there is a conditional lower bound of
$(\poly(n,\allowbreak\log\MaxWeight),\allowbreak m^{1/2-\delta},\allowbreak m^{1-\delta})$
(for any small constant $\delta >0$) for the value of the maximum $s$-$t$ flow
even in unweighted (\ie{} $w \equiv 1)$, undirected graphs based on the OMv
conjecture~\cite{henzinger2015unifying}.
The fastest static algorithm whose running time does not depend on the size of
the largest edge weight computes an optimal solution in $\bigO(nm)$
time~\cite{DBLP:conf/stoc/Orlin13}.
Recently, Chen \etal~\cite{DBLP:conf/focs/ChenGHPS20} gave an $\bigO(\log n
\log \log n)$-approximate fully dynamic algorithm that maintains the maximum
flow value in time $\softO(n^{2/3 + o(1)})$ per update and Goranci
\etal~\cite{DBLP:journals/corr/abs-2005-02369} gave a $n^{o(1)}$-approximate
fully dynamic algorithm to maintain the maximum flow value in time $n^{o(1)}$
worst-case update time and $\bigO(\log^{1/6} n)$ query time.
In the unweighted setting, Jin and Sun~\cite{DBLP:journals/corr/abs-2004-07650}
gave a data structure that can be constructed for any fixed positive integer $c
= (\log n)^{o(1)}$ and that answers for any pair $(s,t)$ of vertices that are
parameters of the query in time $n^{o(1)}$ whether $s$ and $t$ are $c$-edge
connected.

For \emph{global minimum cuts} in the unweighted setting
Thorup and Karger~\cite{DBLP:conf/swat/Thorup00} presented a
$\sqrt{2 + o(1)}$-approximation algorithm to maintain a solution that takes polylogarithmic time per update and query and Thorup~\cite{DBLP:journals/combinatorica/Thorup07} designed a
$(1 + \eps)$-approximate algorithm to maintain a minimum cut in $\softO(\sqrt n)$ time per update
and query.

\paragraph{Experimental Results for Maximum Flow/Minimum Cut}
Kumar and Gupta~\cite{DBLP:journals/jmma/KumarG03} extended the preflow-push
approach~\cite{DBLP:journals/jacm/GoldbergT88} to solve maximum flow in static
graphs to the dynamic setting.
A preflow is a flow under a relaxed conservation constraint in that the excess
of all vertices except $s$ must be non-negative.
Vertices with positive excess are called \emph{active}.
Preflow-push algorithms, also called push-relabel algorithms, use this
relaxed variant of a flow during the construction of a maximum flow along
with distance labels on the vertices.
Generally speaking, they push flow out of active vertices towards vertices with
smaller distance (to $t$) and terminate with a valid flow (i.e., observing
conservation constraints).
In case of an edge insertion or deletion, Kumar and Gupta first identify
affected vertices via forward and backward breadth-first search while observing
and updating distance labels and then follow the scheme of a basic preflow-push
algorithm, however restricted to the set of affected vertices.
The authors evaluated their algorithm only for the incremental setting on a set
of randomly generated instances against the static preflow-push algorithm
in~\cite{DBLP:journals/jacm/GoldbergT88} and found that their algorithm is able
to reduce the number of push and relabel operations significantly as long as
the instances are sparse and the number of affected vertices remains small.

Many important fields of application for the maximum flow/minimum cut problem
stem from computer vision.
In this area, the static algorithm of Boykov and
Kolmogorov~\cite{DBLP:journals/pami/BoykovK04} (\Tool{BK})
is widely used due to its good performance in practice on computer
vision instances and despite its pseudo-polynomial worst-case running time of
$\bigO(nm\cdot\OPT{})$, with $\OPT$ being the value of a maximum flow/minimum
cut.
Interestingly, however, a study by Verma and
Batra~\cite{DBLP:conf/bmvc/VermaB12} shows that its practical superiority only
holds for sparse instances.
\Tool{BK} follows the Ford-Fulkerson method of augmenting flow along
$s$-$t$ paths, but uses two search trees grown from $s$ and $t$, respectively,
to find such paths.
Kohli and Torr~\cite{DBLP:journals/pami/KohliT07,DBLP:series/sci/KohliT10}
extended \Tool{BK} to the fully dynamic setting by updating capacities and flow
upon changes and discuss an optimization that tries to recycle the search
trees.
They experimentally compared their algorithm to repeated executions of the
static algorithm on dynamic instances obtained from video sequences and
achieve a substantial speedup.
They also observed that reusing the search trees leads to longer $s$-$t$ paths,
which affects the update time negatively as the instances undergo many changes.

Goldberg \ea{}~\cite{DBLP:conf/esa/GoldbergHKKTW15} developed \Tool{EIBFS}, a
generalization of their earlier algorithm \Tool{IBFS}, that by contrast also
extends to the dynamic setting in a straightforward manner.
\Tool{IBFS} in turn is a modification of \Tool{BK} that ensures that the two
trees grown from $s$ and $t$ are height-minimal (i.e., BFS trees) and
is closely related to the concept of blocking flows. 
The running time of \Tool{EIBFS} in the static setting and thus the
initialization in the dynamic setting, is $\bigO(mn\log(n^2/m))$ with dynamic
trees or $\bigO(mn^2)$ without.
The algorithm works with a so-called pseudoflow, which observes capacity
constraints, but may violate conservation constraints.
It maintains two vertex-disjoint forests $S$ and $T$, where the roots are
exactly those vertices with a surplus of incoming flow and those with a surplus
of outgoing flow, respectively, and originally only contain $s$ and $t$.
The steps of the algorithm consist in growth steps, where $S$ or $T$ are grown
level-wise, augmentation steps, which occur if a link between the forests has
been established and flow is pushed to a vertex in the other forest and
further on to the root, and adoption steps, where vertices in $T$ with surplus
incoming flow or vertices in $S$ with surplus outgoing flow are either adopted
by a new parent in the same forest or become a root in the other forest.
In case of an update in the dynamic setting, the invariants of the forests are
restored and flow is pushed where possible, followed by alternating
augmentation and adoption steps if necessary.
The authors also mention that resetting the forests every $\bigO(m)$ work such
that they contain only vertices with a surplus outgoing or incoming flow seemed
to be beneficial in practice.
In their experimental evaluation of \Tool{EIBFS} against the algorithm by Kohli
and Torr as well as an altered version thereof and a more naive dynamization of
\Tool{IBFS}, they showed for different dynamic real-world instances from
the field of computer vision that \Tool{EIBFS} is the fastest on eight out of
fourteen instances and relatively robust:
In contrast to its competitors, it always takes at most roughly twice the time
of the fastest algorithm on an instance.
Notably, no algorithm is dominated by another one across all instances.

Zhu \ea{}~\cite{DBLP:journals/tkde/ZhuPSBI16} described a dynamic update
strategy based on augmenting and de-augmenting paths as well as cancelling
cyclic flows.
The latter serves as a preparatory step and only reroutes flow in the network
without increasing or decreasing the total $s$-$t$ flow and is only necessary
in a decremental update operation.
They experimentally evaluated the effectiveness of their algorithm for online
semi-supervised learning, where real-world big data is classified via minimum
cuts, and showed that their algorithm outperforms state-of-the-art stream
classification algorithms.
A very similar algorithm was proposed by Greco
\ea{}~\cite{DBLP:conf/sac/GrecoMPQ17}.
The authors compared it experimentally against \Tool{EIBFS} and the dynamic
\Tool{BK} algorithm by Kohli and Torr as well as a number of the currently
fastest static algorithms.
Their experiments were conducted on a set of instances from computer vision
where equally many edges are randomly chosen to be inserted and deleted,
respectively.
They showed that their algorithm is with one exception always the fastest on
average in performing edge insertions if compared to the average update time of
the competitors, and on half of all instances also in case of edge deletions.
On the remaining instances, the average update time of \Tool{EIBFS} dominated.

\paragraph{Experimental Results for Global Minimum Cuts}
For the global minimum cut problem, Henzinger \etal~\cite{DBLP:journals/corr/abs-2101-05033} implemented an algorithm for large dynamic graphs under both edge insertions and deletions.
For edge insertions, the algorithm uses the approach of
Henzinger~\cite{henzinger1995approximating} and
Goranci~\etal~\cite{goranci2018incremental}, which maintain a compact data
structure of all minimum cuts in a graph and invalidate only the minimum cuts
that are affected by an edge insertion.  For edge deletions,  the algorithms use the push-relabel algorithm of Goldberg and
Tarjan~\cite{goldberg1988new} to certify whether the previous minimum cut is
still a minimum cut. The algorithm outperformed static approaches by up to five orders of magnitude on large graphs.
\ifEnableExtend
\noindent
2003 ``An Incremental Algorithm for the Maximum Flow Problem''~\cite{DBLP:journals/jmma/KumarG03}
2007 ``Dynamic Graph Cuts for Efficient Inference in Markov Random Fields''~\cite{DBLP:journals/pami/KohliT07} \\
2010 ``Dynamic Graph Cuts and Their Applications in Computer Vision''~\cite{DBLP:series/sci/KohliT10}\\
2015 ``Faster and More Dynamic Maximum Flow by Incremental Breadth-First Search''~\cite{DBLP:conf/esa/GoldbergHKKTW15}\\
2016 ``Incremental and Decremental Max-Flow for Online Semi-Supervised Learning''~\cite{DBLP:journals/tkde/ZhuPSBI16}\\
2017 ``Incremental maximum flow computation on evolving networks''~\cite{DBLP:conf/sac/GrecoMPQ17} \Comment{fully dynamic}
\fi

\subsection{Graph Clustering}
Graph clustering is the problem of detecting tightly connected regions of a graph.
More precisely, a clustering $\mathcal{C}$ is a partition of the set of vertices, \ie{} a set of disjoint \emph{clusters} of vertices
  $V_1$,\ldots,$V_k$ such that $V_1\cup\cdots\cup
  V_k=V$. However, $k$ is usually not given in advance and some objective function that models intra-cluster density versus inter-cluster sparsity, is optimized.
It is common knowledge that there is neither a single best strategy nor objective function for graph clustering, which justifies a plethora of existing approaches.
Moreover, most quality indices for graph clusterings have turned out to be
NP-hard to optimize and are rather resilient to effective approximations,
see~\cite{ausiello2012complexity,brandes2008modularity}, \eg{}, allowing only
heuristic approaches for optimization.
There has been a wide-range of algorithms for static graph clustering, the majority are based on the paradigm of intra-cluster density versus inter-cluster sparsity.
For dynamic graphs, there has been a recent survey on the topic of community detection~\cite{DBLP:journals/csur/RossettiC18}. The survey covers features and challenges of dynamic community detection and classifies published approaches. Here we focus on engineering results and extend their survey in that regard with additional references as well as results that appeared in the meantime. A large amount of algorithms in the area optimize for modularity which has been proposed by~\cite{newman2004finding}. The core idea for modularity is to take
coverage, \ie{} the fraction of edges covered by clusters, minus the expected value of the same quantity in a network with the same community divisions, but random connections between the vertices. The commonly used formula is as follows:
$\text{mod}(\mathcal{C})  := \frac{m(\mathcal{C})}{m}- \frac{1}{4m^2} \sum_{C \in \mathcal{C}}\left( \sum_{v \in C} \deg(v) \right)^2$.

\paragraph{Experimental Results}
Miller and Eliassi-Rad~\cite{miller2009continuous} adapted a dynamic extension of Latent Dirichlet Allocation for dynamic graph clustering. Latent Dirichlet Allocation has been originally proposed for modeling text documents, \ie{} the algorithm assumes that a given set of documents can be classified into $k$ topics. This approach has been transferred to graphs~\cite{DBLP:conf/sac/HendersonE09} and was adapted by the authors for dynamic networks.
 Aynaud and Guillaume~\cite{DBLP:conf/wiopt/AynaudG10} tracked communities between successive snapshots of the input network. They first noted that using standard community detection algorithms results in stability issues, \ie{} little modifications of the network can result in wildly different clusterings. Hence, the authors propose a modification of the Louvain method to obtain stable clusterings. This is done by modifying the initialization routine of the Louvain method. By default, the Louvain method starts with each node being in its own clustering. In the modified version of Aynaud and Guillaume, the algorithm keeps the clustering of the previous time step and uses this as a starting point for the Louvain method which results in much more stable clusterings. 
 Bansal \etal~\cite{DBLP:conf/complenet/BansalBP10} also reused the communities from previous time steps. However, their approach is based on greedy agglomeration where two communities are merged at each step to optimize the modularity objective function. The authors improved the efficiency of dynamic graph clustering algorithms by limiting recomputation to regions of the network and merging processes that have been affected by insertion and deletion operations.
G\"orke \etal~\cite{DBLP:journals/jgaa/GorkeHW12} showed that the structure of minimum $s$-$t$-cuts in a graph allows for efficient updates of clusterings. The algorithm builds on partially updating a specific part of a minimum-cut tree and is able to maintain a clustering fulfilling a provable quality guarantee, \ie{} the clusterings computed by the algorithm are guaranteed to yield a certain expansion. To the best of our knowledge, this is the only dynamic graph clustering algorithm that provides such a guarantee.
Later, G\"orke \etal~\cite{DBLP:conf/wea/GorkeMSW10,DBLP:journals/jea/GorkeMSSW13} formally introduced the concept of smoothness to compare consecutive clusterings and provided a portfolio of different update strategies for different types of local and global algorithms. Moreover, their fastest algorithm is guaranteed to run in time $\Theta(\log n)$ per update. Their experimental evaluation indicates that dynamically maintaining a clustering of a dynamic random network saves time and at the same time also yields higher modularity than recomputation from scratch.
Alvari \etal~\cite{DBLP:conf/asunam/AlvariHS14} proposed a dynamic game theory method to tackle the community detection problem in dynamic social networks. Roughly speaking, the authors model the process of community detection as an iterative game performed in a dynamic multiagent environment where each node is an agent who wants to maximize its total utility. In each iteration, an agent can decide to join, switch, leave, or stay in a community. The new utility is then computed by the best outcome of these operations. The authors use neighborhood similarity to measure structural similarity and optimize for modularity. The experimental evaluation is limited to two graphs.
Zakrzweska and Bader~\cite{zakrzweska2015} presented two algorithms that update communities. Their first algorithm is similar to the dynamic greedy agglomeration algorithm by G\"orke \etal~\cite{DBLP:journals/jea/GorkeMSSW13}. The second algorithm is a modification of the first one that runs faster. This first is achieved by more stringent backtracking of merges than  G\"orke \etal~\cite{DBLP:journals/jea/GorkeMSSW13}, \ie{} merges are only undone if the merge has significantly changed the modularity score of the clustering. Moreover, the authors used a faster agglomeration scheme during update operations that uses information about previous merges to speed up contractions.
Recently, Zhuang \etal~\cite{zhuang2019dynamo} proposed the \Tool{DynaMo} algorithm which also is a dynamic algorithm for modularity maximization, however the algorithm processes network changes in batches. 
 \ \\
\noindent 
\ifEnableExtend
2009 ``Continuous time group discovery in dynamic graphs'' Miller Eliassi-Rad~\cite{miller2009continuous} \\
2010 ``Static community detection algorithms for evolving networks'' Aynaud and Guillaume~\cite{DBLP:conf/wiopt/AynaudG10} \\
2010 ``Fast Community Detection for Dynamic Complex Networks'' Bansal \etal~\cite{DBLP:conf/complenet/BansalBP10} \\
2012 ``Dynamic Graph Clustering Using Minimum-Cut Trees'' G\"orke~\cite{DBLP:journals/jgaa/GorkeHW12}\\
2012 ``Modularity-Driven Clustering of Dynamic Graphs''~\cite{DBLP:conf/wea/GorkeMSW10}\\
2013 ``Dynamic graph clustering combining modularity and smoothness''~\cite{DBLP:journals/jea/GorkeMSSW13} \\
2014 ``Community detection in dynamic social networks: {A} game-theoretic approach'' Alvari \etal~\cite{DBLP:conf/asunam/AlvariHS14} \\
2015 ``Fast Incremental Community Detection on Dynamic Graphs'' Zakrzewska and Bader~\cite{zakrzweska2015} \\
2017 ``Local Community Detection in Dynamic Networks'' DiTursi \etal~\cite{DBLP:conf/icdm/DiTursiGB17} \\
2018 ``Community Discovery in Dynamic Networks: {A} Survey''~\cite{DBLP:journals/csur/RossettiC18} \\
2019 ``DynaMo: Dynamic community detection by incrementally maximizing modularity'' Zhuang \etal~\cite{zhuang2019dynamo}

\fi{}
\subsection{Centralities}
We will describe  three popular measures to find central nodes in networks in the fully dynamic setting: Katz centrality, betweenness centrality and closeness centrality.

\emph{Katz centrality} is a centrality metric that measure the relation between vertices by counting weighted walks between them. Formally, the Katz centrality of a vertex $i$ is given as $\textbf{e}_i^T \sum_{k=1}^{\infty} \alpha^{k-1} A^k \textbf{1}$ where $A$ is the adjacency matrix of the graph under consideration, $A^k$ is the $k$th potence of the matrix $A$ (using matrix multiplication), $\textbf{e}_i$ is the $i$th canonical basis vector, $\textbf{e}_i^T$ is the transposed version of $\textbf{e}_i$, $\alpha$ is an input parameter, and $\textbf{1}$ is the vector that contains one at each component.

Given a graph and a vertex $v$ in the graph, the \emph{betweenness centrality} measure is defined to be $c(v) = \sum_{u,w, u\neq w} \frac{\sigma_{u,w}(v)}{\sigma_{u,w}}$, where $\sigma_{u,w}$ is the number of shortest paths between $u$ and $w$ and $\sigma_{u,w}(v)$ is the number of shortest paths between $u$ and $w$ that include $v$. There is also the (less often used) notion of edge betweeness $\tilde c(e) = \sum_{u,w, u\neq w} \frac{\tilde \sigma_{u,w}(e)}{\sigma_{u,w}}$ where $\tilde \sigma_{u,w}(e)$ denotes the number of shortest paths from $u$ to $w$ in $G$ that go though $e$. In both cases, a shortest path is a path with minimum amount of edges. 

Given a graph and a vertex $v$, the \emph{harmonic closeness centrality measure} is defined as $\text{clo}(v) = \sum_{u \in V, u \neq v} \frac{1}{d(u,v)}$ where $d(u,v)$ is the distance between $u$ and $v$.  
Here, distance $d(u,v)$ refers to the number of edges of a shortest path between $u$ and $v$. Roughly speaking, it is the sum of the reciprocal length of the shortest path between the node and all other nodes in the graph. Bavela's definition of \emph{closeness centrality} is similarly $\frac{|V|-1}{\sum_{u\in V} d(v,u)}$.

\paragraph{Theory Results}
The only two theoretical fully dynamic results that we are aware of are due to Pontecorvi and Ramachandran~\cite{DBLP:conf/isaac/PontecorviR15}, who achieve amortized $\bigO{(\nu^{*2}\cdot \log^2 n)}$ update time for betweenness centrality in directed networks where $\nu^*$ bounds the number of distinct edges that lie on shortest paths through any single vertex, and a result due to van den Brand and Nanongkai~\cite{DBLP:conf/focs/BrandN19}, who present a $(1+\epsilon)$-approximate fully-dynamic algorithm for closeness centrality with $\bigO(n^{1.823})$ update time in undirected networks. 
This is an obvious area for future work.

\paragraph{Experimental Results on Katz Centrality}
Nathan and Bader~\cite{DBLP:conf/asunam/NathanB17} were the first to look at the problem in a dynamic setting. 
At that time, static algorithms  mostly used linear algebra-based techniques to compute Katz scores. The dynamic version of their algorithm for undirected graphs incrementally updates the scores by exploiting properties of iterative solvers,~i.e.~Jacobi iterations. Their algorithm achieved speedups of over two orders of magnitude over the simple algorithms that perform static recomputation every time the graph changes.
Later, they improved their algorithm~\cite{DBLP:conf/ppam/NathanB17}
to handle updates  by using an alternative, agglomerative method of calculating Katz scores.  While their static algorithm is already several orders of magnitude faster than typical linear algebra approaches, their dynamic algorithm is also faster than pure static recomputation every time the graph changes.
A drawback of the algorithms by Nathan and Bader is that they are unable to reproduce the exact Katz ranking after dynamic updates. Van der Grinten~\etal{}~\cite{DBLP:conf/esa/GrintenBGBM18} fixed this problem by presenting a dynamic algorithm that iteratively improves upper and lower bounds on the centrality scores. The computed scores are approximate, but the bounds guarantee the correct ranking. The dynamic algorithm improves over the static recomputation of  the Katz rankings as long as the size of the batches in the update sequence is smaller than~\num{10000}. Moreover, the authors are able to deal with undirected and with directed dynamic networks.

\ifEnableExtend
2017 ``Approximating Personalized Katz Centrality in Dynamic Graphs'~\cite{DBLP:conf/ppam/NathanB17} \\
2017 ``A Dynamic Algorithm for Updating Katz Centrality in Graphs''~\cite{DBLP:conf/asunam/NathanB17}                    \\
2018 ``Scalable Katz Ranking Computation in Large Static and Dynamic Graphs''~\cite{DBLP:conf/esa/GrintenBGBM18}\\
             \fi
\paragraph{Experimental Results on Betweenness Centrality} 
Statically computing betweenness centrality involves solving the all-pairs shortest path problem. 
Dynamically maintaining betweenness centrality is challenging as 
the insertion or deletion of a single edge can lead to changes of many shortest paths in the graph.
The \Tool{QUBE} algorithm~\cite{DBLP:conf/www/LeeLPCC12} was the first to provide a non-trivial update routine for undirected graphs. The key idea is to perform the betweenness computation on a reduced set of vertices, \ie{} the algorithm first finds vertices whose centrality index might have changed. Betweenness centrality is then only computed on the first set of vertices. However, \Tool{QUBE} is limited to the insertion and deletion of non-bridge edges.
Lee \etal~\cite{DBLP:journals/isci/LeeCC16} extended the \Tool{QUBE} algorithm~\cite{DBLP:conf/www/LeeLPCC12} to be able to insert and delete non-bridge edges. Moreover, the authors reduced the number of shortest paths that need to be recomputed and thus gained additional speedups over \Tool{QUBE}.
Kourtellis \etal~\cite{DBLP:conf/icde/KourtellisMB16,DBLP:journals/tkde/KourtellisMB15} contributed an algorithm that maintains both vertex and edge betweenness centrality (for directed as well as undirected networks). Their algorithm needs less space than the algorithm by Green \etal{}~\cite{DBLP:conf/socialcom/GreenMB12} as it avoids storing predecessor lists. Their method can be parallelized and runs on top of parallel data processing engines such as Hadoop. 
Bergamini \etal~\cite{DBLP:conf/alenex/BergaminiMS15} presented an incremental \emph{approximation} algorithm for the problem which is based on the first theory result that is asymptotically faster than recomputing everything from scratch due to Nasre~\etal~\cite{DBLP:conf/mfcs/NasrePR14}. The algorithm works for directed as well as for undirected networks. As a building block of their algorithm, the authors used an asymptotically faster algorithm for the dynamic single-source shortest path problem and additionally sample shortest paths. Experiments indicate that the algorithm can be up to four orders of magnitude faster compared to restarting the static approximation algorithm by Riondato and Kornaropoulos~\cite{DBLP:conf/wsdm/RiondatoK14}.
In the same year, the authors extended their algorithm to become a fully dynamic approximation algorithm for the problem~\cite{DBLP:conf/esa/BergaminiM15,DBLP:journals/im/BergaminiM16}. In addition to dynamic single-source shortest paths, the authors also employed an approximation of the vertex diameter that is needed to compute the number of shortest paths that need to be sampled 
as a function of a given error guarantee that should be achieved. 
Hayashi~\etal~\cite{DBLP:journals/pvldb/HayashiAY15} provided a fully dynamic approximation algorithm for directed networks that is also based on sampling. In contrast to Bergamini \etal{}~\cite{DBLP:conf/alenex/BergaminiMS15,DBLP:conf/esa/BergaminiM15,DBLP:journals/im/BergaminiM16}, which samples shortest paths between randomly selected vertices, the authors save all the paths between each sampled pair of vertices. Moreover, the shortest paths are represented in a data structure called hypergraph sketch. To further reduce the running time when handling unreachable pairs, the authors maintain a reachability index. 
Gil-Pons~\cite{DBLP:conf/ciarp/Pons18} focused on exact betweenness in incremental directed graphs. The author presented a space-efficient algorithm with linear space complexity.
Lastly, Chehreghani \etal~\cite{DBLP:conf/bigdataconf/ChehreghaniBA18a} focused on the special case in which the betweenness of a single node has to be maintained under updates (vertex/edge insertion/deletion) in directed graphs. 

\ifEnableExtend
2012 ``{QUBE:} a quick algorithm for updating betweenness centrality''~\cite{DBLP:conf/www/LeeLPCC12} \\
2012 ``A Fast Algorithm for Streaming Betweenness Centrality''~\cite{DBLP:conf/socialcom/GreenMB12}\\

2013  ``Incremental algorithm for updating betweenness centrality in dynamically growing networks''~\cite{DBLP:conf/asunam/KasWCC13} \\
2015 ``Fully-Dynamic Approximation of Betweenness Centrality''~\cite{DBLP:conf/esa/BergaminiM15} \\
2015 ``Fully Dynamic Betweenness Centrality Maintenance on Massive Networks''~\cite{DBLP:journals/pvldb/HayashiAY15} \\
2016 ``Approximating Betweenness Centrality in Fully Dynamic Networks''~\cite{DBLP:journals/im/BergaminiM16}      \\
2016 ``Scalable online betweenness centrality in evolving graphs''~\cite{DBLP:conf/icde/KourtellisMB16,DBLP:journals/tkde/KourtellisMB15} \\
2016 ``Efficient algorithms for updating betweenness centrality in fully dynamic graphs''~\cite{DBLP:journals/isci/LeeCC16} \\
2017   ``Faster Betweenness Centrality Updates in Evolving Networks''~\cite{DBLP:conf/wea/BergaminiMOS17} \\
2018 ``Space Efficient Incremental Betweenness Algorithm for Directed Graphs''~\cite{DBLP:conf/ciarp/Pons18} \\
2018 ``Scaling Betweenness Centrality in Dynamic Graphs''~\cite{DBLP:conf/hpec/TripathyG18} \\
2018 ``DyBED: An Efficient Algorithm for Updating Betweenness Centrality in Directed Dynamic Graphs''~\cite{DBLP:conf/bigdataconf/ChehreghaniBA18a}\\ -- only a single node is fixed here
\fi

\paragraph{Experimental Results on Closeness Centrality} 
Kas \etal~\cite{DBLP:conf/asunam/KasCC13} were the first to give a \emph{fully dynamic} algorithm for the problem (directed/undirected).
As computing closeness centrality depends on the all-pairs shortest path problem, the authors extended an existing dynamic all-pairs shortest path algorithm~\cite{ramalingam1991computational} for their problem. 
As the algorithm stores pairwise distances between nodes it has quadratic memory requirement.
Sariyuce \etal~\cite{DBLP:conf/bigdataconf/SariyuceKSC13} provided an algorithm that can handle insertions and deletions in undirected graphs. In contrast to Kas \etal~\cite{DBLP:conf/asunam/KasCC13}, the authors used static single-source shortest paths from each vertex. The algorithm does not need to store pairwise distances and hence requires only a linear amount of memory. Moreover, the authors observed that in scale-free networks the diameter grows proportional to the logarithm of the number of nodes, \ie{} the diameter is typically small. When the graph is modified with minor updates, the diameter also tends to stay small. This can be used to limit the number of vertices that need to be updated. In particular, the authors showed that recomputation of closeness can be skipped for vertices $s$ such that $|d(s,u)-d(s,v)| = 1$ where $u$, $v$ are the endpoints of the newly inserted edge. Lastly, the authors used data reduction rules to filter vertices, \ie{} real-life networks can contain nodes that have the same or similar neighborhood structure that can be merged. Later, Sariyuce~\etal~\cite{DBLP:conf/cluster/SariyuceSKC13,DBLP:journals/pc/SariyuceSKC15} proposed a distributed memory-parallel algorithm for the problem in undirected dynamic graphs. 
Yen \etal~\cite{DBLP:conf/icdm/YenYC13} proposed the fully dynamic algorithm \Tool{CENDY} which can reduce the number of internal updates to a few single-source shortest path computations necessary by using breadth-first searches. The authors introduce the notion of augmented rooted BFS trees, which are rooted BFS trees plus two types of edges that indicate if two vertices are on the same level of the BFS tree or if they are on different levels of the BFS tree. The main idea is that given an augmented rooted BFS tree of an unweighted network, edges that are inserted or deleted within the same level of the tree do not change the distances from the root to all other vertices.
Putman \etal~\cite{DBLP:conf/asunam/PutmanBT19} provided a faster algorithm for fully dynamic harmonic closeness in directed networks.
The authors also used a filtering method to heavily reduce the number of computations for each incremental update. The filtering method is an extension of level-based filtering to directed and weighted networks.
The dynamic algorithm by Shao \etal~\cite{DBLP:conf/dasfaa/ShaoGGWLY20} maintains closeness centrality by efficiently detecting all affected shortest paths based on articulation points. The main observation is that a graph can be divided into a series of biconnected components which are connected by articulation points -- the distances between two arbitrary vertices in the graph can be expressed as multiple distances between different biconnected components.

Bisenius \etal~\cite{DBLP:conf/alenex/BiseniusBAM18} contributed an algorithm to maintain top-$k$ harmonic closeness in fully dynamic graphs for undirected and directed networks. The algorithm is not required to compute closeness centrality for the initial graph and the memory footprint of their algorithm is linear. Their algorithm also tries to skip recomputations of vertices that are unaffected by the modifications of the graph by running breadth-first searches.
Crescenzi \etal~\cite{DBLP:journals/algorithms/CrescenziMM20} gave a fully dynamic \emph{approximation} algorithm for top-$k$ harmonic closeness for undirected as well as directed neworks. The algorithm is based on sampling paths and a backward dynamic breadth-first search~algorithm.
\ifEnableExtend
2013 ``Incremental closeness centrality for dynamically changing social networks''~\cite{DBLP:conf/asunam/KasCC13} \\

2013 ``Incremental algorithms for closeness centrality''~\cite{DBLP:conf/bigdataconf/SariyuceKSC13} \\
2013 ``{STREAMER:} {A} distributed framework for incremental closeness centrality''~\cite{DBLP:conf/cluster/SariyuceSKC13},

2013 ``An Efficient Approach to Updating Closeness Centrality and Average Path Length in Dynamic Networks'~\cite{DBLP:conf/icdm/YenYC13} \\
2015 ``Incremental closeness centrality in distributed memory''~\cite{DBLP:journals/pc/SariyuceSKC15} \\

2018 ``Computing Top-\emph{k} Closeness Centrality in Fully-dynamic Graphs''~\cite{DBLP:conf/alenex/BiseniusBAM18} \\
2019 ``Fast Incremental Computation of Harmonic Closeness Centrality in Directed Weighted Networks''~\cite{DBLP:conf/asunam/PutmanBT19} \\
2020 ``Efficient Closeness Centrality Computation for Dynamic Graphs''~\cite{DBLP:conf/dasfaa/ShaoGGWLY20} \\

2020 ``Finding Top-k Nodes for Temporal Closeness in Large Temporal Graphs''~\cite{DBLP:journals/algorithms/CrescenziMM20}\\ 

\fi
\subsection{Graph Partitioning}
\label{sss:gp}
Typically the graph partitioning problem asks for a partition of a graph into $k$ blocks of about equal size such that there are few edges between them.
More formally, we are given a graph $G=(V, E)$ with vertex weights $c:V \rightarrow \mathbb{R}_{>0}$ and edge  
weights $\omega:E \rightarrow \mathbb{R}_{>0}$.
We extend $c$ and $\omega$ to sets in the natural way, i.e., $c(U) :=\sum_{v\in U} c(v)$ and $\omega(F) :=\sum_{e \in F} \omega(e)$.
The graph partitioning problems is looking for disjoint \emph{blocks} of vertices $V_1$,\ldots,$V_k$ that partition $V$, i.e.,
$V_1\cup\cdots\cup V_k=V$. A \emph{balancing constraint} demands that all blocks
have weight $c(V_i)\leq (1+\epsilon)\lceil\frac{c(V)}{k}\rceil$
for some imbalance parameter $\epsilon$.
The most used objective is to minimize the total \emph{cut} which is defined as $\omega(E\cap\bigcup_{i<j}V_i\times V_j)$.
The problem is known to be NP-hard and no constant-factor approximation algorithms exist. Thus heuristic algorithms are mostly used in practice. Dynamic graph partitioning algorithms are also known in the community as \emph{repartitioning algorithms.} As the problem is typically not solved to optimality in practice, repartitioning involves a tradeoff between the \emph{quality}, \ie{} the number of edges in different sets of the partitioning,  and the amount of vertices that need to change their block as they cause communication when physically moved between the processors as the partition is adopted.
The latter is especially important when graph partitioning is used in adaptive numerical simulations. In these simulations, the main goal is to partition a model of computation and communication in which nodes model computation and edges model communication. The blocks of the partition are then fixed to a specific processing element. When the dynamic graph partitioning algorithm decides to change the blocks due to changes in the graph topology, nodes that are moved to a different block create communication in the simulation system as the underlying data needs to be moved between the corresponding processors.

\paragraph{Experimental Results}
Hendrikson \etal~\cite{DBLP:conf/hicss/HendricksonLD96} tackled the repartitioning problem by introducing $k$ virtual vertices. Each of the virtual vertices is connected to all nodes of its corresponding block. The edges get a weight $\alpha$  which is proportional to the migration cost of a vertex and the vertex weights of the virtual vertices are set to zero. Then an updated partition can be computed using a static partitioning algorithm since the model accounts for migration costs and edge cut size at the same time.

Schloegel \etal~\cite{DBLP:journals/jpdc/SchloegelKK97}  presented heuristics to control the tradeoff between edge-cut size and vertex migration costs. The most simple algorithm is to compute a completely new partition and then determine a mapping between the blocks of the old and the new partition that minimizes migration. The more sophisticated algorithm  of~\cite{DBLP:journals/jpdc/SchloegelKK97} is a multilevel algorithm based on a simple process, 
\ie{} nodes are moved from blocks that contain too many vertices to blocks that contain not enough vertices. However, this approach often yields partitions that cut a large number of edges. The result has been improved later by combining the two approaches in the parallel partitioning tool \Tool{ParMetis}~\cite{DBLP:conf/sc/SchloegelKK00}. Schloegel \etal~\cite{DBLP:journals/concurrency/SchloegelKK02} later extended their algorithm to be able to handle multiple balance constraints.
Hu and Blake~\cite{HU1999417} noted that diffusion processes can suffer from slow convergence and improved the performance of diffusion through the use of Chebyshev polynomials. More precisely, the diffusion process in their paper is a directed diffusion  that computes a diffusion solution by solving a so-called head conduction equation while minimizing the data movement.
Walshaw \etal~\cite{DBLP:journals/jpdc/WalshawCE97} integrated a repartitioning algorithm into their parallel (meanwhile uncontinued) tool \Tool{Jostle}. The algorithm is a directed diffusion process based on the solver proposed by Hu and Blake~\cite{HU1999417}. Rotaru and N\"ageli~\cite{rotaru2004dynamic} extended previous diffusion-based algorithms to be able to handle heterogeneous systems. 
These approaches, however, have certain weaknesses: For example, in numerical applications the maximum number of boundary nodes of a block is often a better estimate of the occurring communication in the simulation than the number of cut edges. 
Meyerhenke and Gehweiler~\cite{meyerhenke2009dynamic,DBLP:conf/dagstuhl/MeyerhenkeG10} explored a disturbed diffusion process that is able to overcome some of the issues of the  previous approaches.
To do so, Meyerhenke adapted \Tool{DIBAP}, a previously developed algorithm that aims at computing well-shaped partitions.
A diffusion process is called disturbed if its convergence state does not result in a balanced distribution. These processes can be helpful to find densely connected regions in the graph. 

There has been also work that tackles slightly different problem formulations. Kiefer \etal~\cite{DBLP:conf/europar/KieferHL16} noted that performance in applications usually does not scale linearly with the amount of work
per block due to contention on different compute components. Their algorithm uses a simplified penalized resource consumption model. Roughly speaking, the authors introduced a penalized block weight 
and modified the graph partitioning problem accordingly. More precisely, a positive, monotonically increasing penalty function $p$ is used to penalize the weight of a block  based on the partition cardinality.
Vaquero \etal~\cite{DBLP:conf/icdcs/VaqueroCLM14} looked at the problem for distributed graph processing systems. Their approach is based on iterative vertex migration based on label propagation. More precisely, a vertex has a list of candidate blocks where the highest number of its neighbors are located. However, initial partitions are computed using hashing which does not yield high quality partitions since it completely ignores the structure of the graph. The authors did not compare their work against other state-of-the art repartitioning algorithms, so it is unclear how well the algorithm performs compared to other algorithms. 
Xu~\etal~\cite{xu2014loggp} and Nicoara \etal~\cite{nicoara2015hermes} also presented  dynamic algorithms specifically designed for graph processing systems. Other approaches have focused on the edge partitioning problem~\cite{DBLP:conf/ideas/SakouhiAGSM16,DBLP:journals/pvldb/HuangA16,fan2020incrementalization} or the special case of road networks~\cite{buchhold2020fast}.
\ifEnableExtend
1997 ``Multilevel Diffusion Schemes for Repartitioning of Adaptive Meshes''~\cite{DBLP:journals/jpdc/SchloegelKK97} \\
1997 ``Parallel dynamic graph partitioning for adaptive unstructured meshes''~\cite{DBLP:journals/jpdc/WalshawCE97} -- parallel, not included\\ 
1998 ``Dynamic repartitioning of adaptively refined meshes''~\cite{schloegel1998dynamic}  -- too old           \\
1999 ``An improved diffusion algorithm for dynamic load balancing''~\cite{HU1999417} \\
2002 ``Parallel static and dynamic multi-constraint graph partitioning''~\cite{DBLP:journals/concurrency/SchloegelKK02} \\
2004 ``Dynamic load balancing by diffusion in heterogeneous systems''~\cite{rotaru2004dynamic}\\
2006 ``Part. and dynamic load balancing for the numerical solution of partial differential equations''~\cite{teresco2006partitioning}\\
2007 ``Hypergraph-based dynamic load balancing for adaptive scientific computations''~\cite{catalyurek2007hypergraph} -- hypergraphs not graphs\\
2009 ``A repartitioning hypergraph model for dynamic load balancing''~\cite{catalyurek2009repartitioning}-- hypergraphs not graphs\\
2009 ``Dyn. load balancing for par. numerical simulations based on repartitioning with dist. diffusion''~\cite{meyerhenke2009dynamic}\\
2010 ``On dynamic graph partitioning and graph clustering using diffusion'~\cite{DBLP:conf/dagstuhl/MeyerhenkeG10}\\

2013 ``Dynamic partitioning of big hierarchical graphs''~\cite{DBLP:conf/vldb/SpyropoulosK13} -- no citations\\
2014 ``Adaptive partitioning for large-scale dynamic graphs''~\cite{DBLP:conf/icdcs/VaqueroCLM14}\\
2014 ``LogGP: a log-based dynamic graph partitioning method'~\cite{xu2014loggp} \\
2015  ``Hermes: Dynamic Partitioning for Distributed Social Network Graph Databases''~\cite{nicoara2015hermes} \\
2016 ``Dynamicdfep: A distributed edge partitioning approach for large dynamic graphs''~\cite{DBLP:conf/ideas/SakouhiAGSM16} -- edge partitioning, excluded\\
2016 ``LEOPARD: lightweight edge-oriented partitioning and replication for dynamic graphs''~\cite{DBLP:journals/pvldb/HuangA16} -- edge partitioning, excluded\\
2016 ``Penalized graph partitioning for static and dynamic load balancing''~\cite{DBLP:conf/europar/KieferHL16} \\
2020 ``Fast and Stable Repartitioning of Road Networks''~\cite{buchhold2020fast} \\
2020 ``Incrementalization of graph partitioning algorithms''~\cite{fan2020incrementalization}\\
\fi

\section{Dynamic Graph Systems}
The methodology of the previous two sections is to engineer algorithms for specific dynamic graph problems. In contrast to this, there are also approaches that try to engineer dynamic graph systems that can be applied to a wide range of dynamic graph problems. Alberts \etal~\cite{alberts1998software} started this effort and presented a software library of dynamic graph algorithms. The library is written in C++ and is an extension of the well known \Tool{LEDA} library of efficient data types and algorithms. The library contains algorithms for connectivity, spanning trees, single-source shortest paths and transitive closure.\\ 
A decade later Weigert \etal~\cite{weigert2011mining} presented a system that is able to deal with dynamic distributed graphs, \ie{} in settings in which a graph is too large for the memory of a single machine and, thus, needs to be distributed over multiple machines. A user can implement a query function to implement graph queries. Based on their experiments, the system appears to scale well to large distributed graphs.
Ediger \etal~\cite{DBLP:conf/hpec/EdigerMRB12} engineered \Tool{STINGER} which is an acronym for Spatio-Temporal Interaction Networks and Graphs Extensible Representation. \Tool{STINGER} provides fast insertions, deletions, and updates on semantic graphs that have a skewed degree distribution. The authors showed in their experiments that the system can handle 3 million updates per second on a scale-free graph with 537 million edges on a Cray XMT machine. The authors already implemented a variety of algorithms on \Tool{STINGER} including community detection, $k$-core extraction, and many more.
Later, Feng \etal~\cite{feng2015distinger} presented \Tool{DISTINGER} which has the same goals as \Tool{STINGER}, but focuses on the distributed memory case, i.e. the authors presented a distributed graph representation.
Vaquero \etal~\cite{DBLP:journals/corr/VaqueroCLM13} presented a dynamic graph processing system that uses adaptive partitioning to update the graph distribution over the processors over time. This speeds up queries as a better graph distribution significantly reduces communication overhead. Experiments showed that the repartitioning heuristic (also explained in Section~\ref{sss:gp}) improves computation performance in their system up to \SI{50}{\percent} for an algorithm that computes the estimated diameter in a graph.
Sengupta \etal~\cite{sengupta2016graphin} introduced a dynamic graph analytics framework called \Tool{GraphIn}. Part of \Tool{GraphIn} is a new programming model based on the gather-apply-scatter programming paradigm that allows users to implement a wide range of graph algorithms that run in parallel. 
Compared to \Tool{STINGER}, the authors reported a \num{6.6}-fold speedup.
Iwabuchi \etal~\cite{iwabuchi2016towards} presented an even larger speedup over \Tool{STINGER}. Their dynamic graph data store is, like \Tool{STINGER}, designed for scale-free networks. The system uses compact hash tables with high data locality. In their experiments, their system called \Tool{DegAwareRHH}, is a factor \num{206.5} faster than~\Tool{STINGER}.

Another line of research focuses on graph analytic frameworks and data structures for GPUs. Green and Bader~\cite{feng2015distinger} presented \Tool{cuSTINGER}, which is a GPU extension of STINGER and targets NVIDIA's CUDA supported GPUs. One drawback of \Tool{cuSTINGER} is that the system has to perform restarts after a large number of edge updates. Busato \etal \cite{DBLP:conf/hpec/BusatoGBB18} fixed this issue in their system, called \Tool{Hornet}, and, thus, outperform \Tool{cuSTINGER}. Moreover, \Tool{Hornet} uses a factor of 5 to 10 less memory than \Tool{cuSTINGER}.
In contrast to previous approaches, \Tool{faimGraph} due to Winter \etal~\cite{DBLP:conf/sc/WinterMZSS18} is able to deal with a changing number of vertices.
Awad \etal~\cite{DBLP:conf/ipps/AwadAPO20} noted that the experiments performed
by Busato \etal are missing true dynamism that is expected in real world
scenarios and proposed a dynamic graph structure that uses one hash table per
vertex to store adjacency lists. The system achieves speedups between \numrange{5.8}{14.8}
compared to \Tool{Hornet} and \numrange{3.4}{5.4} compared to \Tool{faimGraph} for batched edge
insertions (and slightly smaller speedups for batched edge deletions). The
algorithm also supports vertex deletions, as does \Tool{faimGraph}.

 \ifEnableExtend
1998 ``A software library of dynamic graph algorithms''~\cite{alberts1998software} \\
2011 ``Mining large distributed log data in near real time''~\cite{weigert2011mining} \\
2012 ``{STINGER:} High performance data structure for streaming graphs'~\cite{DBLP:conf/hpec/EdigerMRB12} \\
2013 ``xDGP: {A} Dynamic Graph Processing System with Adaptive Partitioning''~\cite{DBLP:journals/corr/VaqueroCLM13} \\
2015 ``DISTINGER: A distributed graph data structure for massive dynamic graph processing''~\cite{feng2015distinger} \\
2016 ``cuSTINGER: Supporting dynamic graph algorithms for GPUs''~\cite{green2016custinger} \\
2016 ``Graphin: An online high performance incremental graph processing framework''~\cite{sengupta2016graphin} \\
2016 ``Towards a distributed large-scale dynamic graph data store''~\cite{iwabuchi2016towards} \\
2017 ``Accelerating Dynamic Graph Analytics on GPUs''~\cite{DBLP:journals/pvldb/ShaLHT17} \\
2018 ``Hornet: An efficient data structure for dynamic sparse graphs and matrices on gpus''~\cite{busato2018hornet} \\
2019 ``Graphtinker: A high performance data structure for dynamic graph processing''~\cite{jaiyeoba2019graphtinker} \\
2020 ``Dynamic Graphs on the {GPU}''~\cite{DBLP:conf/ipps/AwadAPO20}
 \fi{}

\section{Methodology}
Currently there is a limited amount of real-world \emph{fully} dynamic networks publicly available. There are repositories that feature a lot of  real-world insertions only instances such as SNAP\footnote{\url{https://snap.stanford.edu/}} and KONECT\footnote{\url{http://konect.cc/}}. However, since the fully dynamic instances are rarely available at the moment, we start a new graph repository that provides fully dynamic graph instances\footnote{\url{https://DynGraphLab.github.io}}. Currently, there is also very limited work on dynamic graph generators. A generator for clustered dynamic random networks has been proposed by G\"orke \etal~\cite{DBLP:conf/medalg/GorkeKSSW12}. Another approach is due to Sengupta~\cite{DBLP:conf/icdm/SenguptaHW17} to generate networks for dynamic overlapping communities in networks. A generative model for dynamic networks with community structure can be found in~\cite{maBecker}. This is a widely open topic for future work, both in terms of oblivious adversaries as well as adaptive adversaries.
To still be able to evaluate fully dynamic algorithms in practice, research uses a wide range of models at the moment to turn static networks into dynamic ones. We give a brief overview over the most important ones.
In \emph{undo-based} approaches, edges of a static network are inserted in some order until all edges are inserted. In the end, \SI[parse-numbers=false]{x}{\percent} of the last insertions are undone. The intuition here is that one wants undo changes that happened to a network and to recreate a previous state of the data structure.
In \emph{window-based} approaches, edges are inserted and have a predefined lifetime. That means an edge is deleted after a given number $d$ of new edges have been inserted. 
In \emph{remove and add} based approaches, a small fraction of random edges from a static network is removed and later on reinserted. In practice, researchers use a single edge as well as whole batches of edges.
In \emph{morphing-based} approaches, one takes two related networks and creates a sequence of edge updates such that the second network obtained after the update sequence has been applied to the~first~network.

\section{Final Remarks}
Traditionally, algorithms are designed using simple models.
In turn, performance guarantees are provable for all possible inputs.
For researchers working in algorithm theory, however, implementing an algorithm
is only part of the application development.
This causes a growing gap between theory
and practice.
For dynamic algorithms, this works in both ways. On the one hand, there is a large body of theoretical work on efficient dynamic graph algorithms that received very little attention from practitioners (\eg{} spanning trees or diameter). On the other hand, there is a range of problems with a large amount of practical work, yet theoretical foundations seem to be (somewhat) missing (\eg{} centralities, graph partitioning/clustering). For some problems theoretical and practical work exists to a large extend, but appear to be disconnected (\eg{} motif counting, maximum flows). Lastly, there are also a range of problems in which ideas from theory have been picked up and evaluated by practitioners (\eg{} reachability, matching, shortest paths). \\

\textbf{Acknowledgements.}
\erclogowrapped{4\baselineskip}
This project has received funding from the
European Research Council (ERC) under the European Union's Horizon 2020
research and innovation programme (Grant agreement No.\ 101019564
``The Design of Modern Fully Dynamic Data Structures (MoDynStruct)''
and from the
Austrian Science Fund (FWF) project ``Fast Algorithms for a Reactive Network
Layer (ReactNet)'', P~33775-N, with additional funding from the \textit{netidee SCIENCE
Stiftung}, 2020--2024.
Moreover, we have been partially supported by DFG grant SCHU 2567/1-2.
\bibliographystyle{plainnat}
\bibliography{references.bib}

\end{document}